\documentclass[fleqn,usenatbib]{mnras}

\usepackage{newtxtext,newtxmath,xcolor,dsfont}
\usepackage{hyperref}
\usepackage{appendix}
\usepackage{soul}
\hypersetup{
    colorlinks=true,
    linkcolor=blue,
    filecolor=magenta,      
    urlcolor=cyan,
    pdftitle={Movie},
    pdfpagemode=FullScreen,
    }

\DeclareRobustCommand{\VAN}[3]{#2}
\let\VANthebibliography\thebibliography
\def\thebibliography{\DeclareRobustCommand{\VAN}[3]{##3}\VANthebibliography}

\newcommand{\degree}{^{\circ}}
\definecolor{DarkRed}{rgb}{0.35, 0.035, 0.035}
\definecolor{DarkBG}{rgb}{0.01 0.1 0.6}

\newcommand{\rev}[1]{\textcolor{black} { {#1}}}
\newcommand{\revv}[1]{\textcolor{black} { {#1}}}
\newcommand{\rrev}[1]{\textcolor{black} { {#1}}}

\newcommand{\PS}{$\mathds{P}$}
\newcommand{\radmm}{~ rad~m$^{-2}$}
\newcommand{\vb}{{\bf \emph{B}}~ }

\usepackage{graphicx}	
\usepackage{amsmath}	

\title[\rev{Pseudo-3D Faraday visualization}]{\rev{Pseudo-3D visualization of Faraday structure in polarized radio sources:  methods, science use cases, and development priorities}}
\author[L. Rudnick, C. Anderson, W. D. Cotton et al.]{
Lawrence Rudnick,$^{1}$\thanks{E-mail: larry@umn.edu (LR)}
C. Anderson,$^{2}$, W. D. Cotton$^{3}$, A. Pasetto$^4$, E. Alexander$^5$, M. Tahani $^6$\\
$^{1}$Minnesota Institute for Astrophysics, University of Minnesota, 116 Church St. SE, Minneapolis, MN 55455, USA\\
$^{2}$Research School of Astronomy and Astrophysics, Australian National University, Canberra, AUS\\
$^{3}$National Radio Astronomy Observatory, Charlottesville, VA, USA\\
$^4$ Institute for Radio Astronomy and Astrophysics, National Autonomous University of Mexico,Michoacán, México \\
$^5$ Jodrell Bank Centre for Astrophysics, Department of Physics and Astronomy, University of Manchester, Manchester, UK\\
$^6$ Kavli Institute for Particle Astrophysics and Cosmology, Stanford University, Stanford, CA, USA\\
}

\date{ \today}
\pubyear{2024}

\begin{document}
\label{firstpage}
\pagerange{\pageref{firstpage}--\pageref{lastpage}}
\maketitle

\begin{abstract}
We introduce the construction of polarized intensity cubes $\mathds{P}$(\textrm{RA}, \textrm{Dec}, $\Phi$) \rev{and their visualization as movies}, as a powerful technique for interpreting Faraday structure.  $\mathds{P}$ is constructed from  maps of peak polarized intensity P(\textrm{RA}, \textrm{Dec}) with their corresponding Faraday depth maps $\Phi$(\textrm{RA}, \textrm{Dec}).  \rev{We illustrate the extensive scientific potential  of such visualizations with a variety of science use cases from}   ASKAP and MeerKAT, \rev{presenting models that are consistent with the data but not necessarily unique.} We demonstrate how one can, in principle,  \rev{distinguish between cube structures which originate} from  unrelated foreground screens from those due to \rev{magnetized plasmas} local to the \rev{emitting} source. \rev{Other science use cases illustrate how variations in the local $n_e$  $B$,  and line-of-sight distance to the synchrotron emitting regions can be distinguished using Faraday rotation. We show, for the first time, how the line-of-sight orientation of AGN jets can be determined. We also examine the case of M87 to show how internal jet magnetic field configurations can be identified, and extend earlier results.} 
We recommend using this technique to re-evaluate all previous analyses of polarized sources that are well-resolved both spatially and in Faraday depth. \rev{Recognizing the subjective nature of interpretations at this early stage, } we also highlight the need and utility for further scientific and technical developments.
\end{abstract}

\begin{keywords}
Galaxies: active -- Galaxies: magnetic fields -- magnetic fields -- polarization -- Methods: observational
\end{keywords}

\section{Introduction}

\rev{Maps of the peak Faraday depth $\Phi$\footnote{\rev{This is equivalent to the ``rotation measure" (RM), when there is a single peak in the Faraday spectrum, or when  \rev{$\Phi$} is calculated by fitting a slope to the observed variation of polarization angle vs. wavelength-squared}} in a Faraday spectrum provide information on the magnetized thermal medium along each line of sight to the synchrotron emission region. When there is a single dominant Faraday depth $\Phi$ in each observing beam, $\Phi = \int{n_e(\bf{s})~ \bf{B(s)}\cdot d\bf{s}}$, where $\bf{B}$ is the magnetic field, and  \rrev{${n_e}$} is the electron density along the line of sight $\bf{s}$. These maps thus contain information about foreground magnetized thermal plasmas that are \emph{unrelated} to the source, such as the Milky Way, as well as about the medium local to, or even mixed with the synchrotron emitting plasma.  This paper introduces a powerful visualization technique to distinguish between these unrelated foregrounds and local media, as well as to probe the physical conditions when the Faraday medium is local to the source. } 

\begin{figure}
 \centering
    \includegraphics[width=2.in]{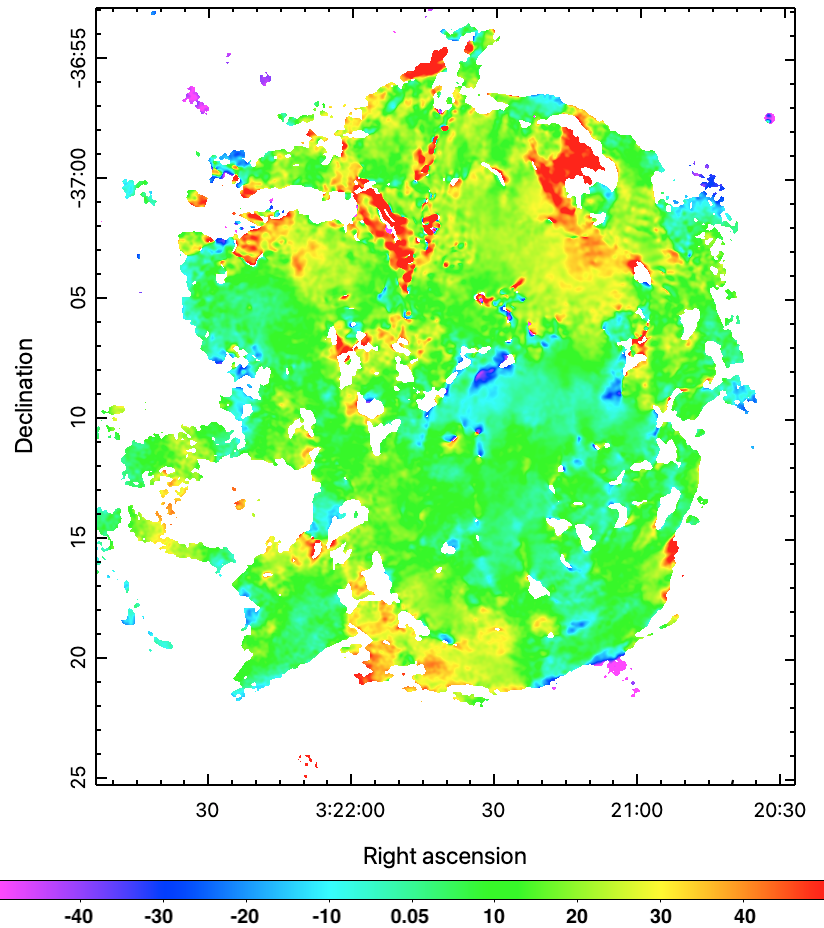}
    \caption{Peak Faraday depth image of Fornax A, western lobe, from \citet{2021PASA...38...20A}.}
    \label{fig:ForAW_RMB}
\end{figure}

\rev{A variety of methods to display the variations in $\Phi$  for extended sources have been used.  The earliest results were presented as one-dimensional plots of $\Phi$ along a source \citep{1975AJ.....80..559S, 1979A&AS...36..173H}. Displaying 2D $\Phi$ distributions was more challenging, and grids of numbers or symbols were used, along with contours \citep{1980ApJ...236..761D,1984ApJS...54..291P}.  Greyscale maps of RM  were an improvement on this \citep{1986A&A...156..234L}, and finally false-color images were introduced \citep{2006MNRAS.368...48L} and today remain the almost exclusive type of display \citep[see Fig. \ref{fig:ForAW_RMB} here, and ][]{Knowles2021, 2023ApJ...955...16B,2022ApJ...937...45A}. None of these methods facilitate finding correlations of $\Phi$ structures with the synchrotron structures, or even revealing spatial patterns in $\Phi$, unless they are quite obvious, such as the bands reported by \cite{LG2}.}

\rev{ A significant advance came with the development of 
Faraday Synthesis \citep{Brentjens2005}, which allowed the detection of multiple Faraday depths along the same line of sight. A cube $\mathds{F}$(RA, Dec, $\Phi$) is produced, where the $\Phi$ axis is the Faraday spectrum at each position. \cite{2011A&A...526A...9B} used a $\mathds{F}$(RA, Dec, $\Phi$) cube of the Perseus cluster of galaxies    to display results as a series of fixed $\Phi$(RA,Dec) frames. 
A significant new type of display,  a 2D image of $\Phi$ vs. a spatial dimension, was introduced by \cite{2011A&A...525A.104P} but has not been well-utilized in the literature. 
The richness of the information that could be extracted from these cubes is apparent in the animations of radio structures in Abell~194 \citep{A194},  closely related to the technique introduced here.  }

\rev{In order to understand the need for more powerful forms of visualization, we briefly review how maps of Faraday depth have been used. They have yielded extensive information on}
the structure and strength of magnetic fields in foreground screens, unrelated to the emitting source, e.g., the magnetic structure of the Milky Way \citep{1979Natur.279..115S,2009ApJ...702.1230T,2015A&A...575A.118O,2022A&A...657A..43H}, nearby galaxies \citep{1998A&A...335.1117H} and the intervening intracluster medium \citep{2003ASPC..301..501G}. 
\rev{The new visualization technique introduced here does \emph{not} add any value in these cases of unrelated foregrounds, so we turn now to Faraday structures local to the emitting source.}

\rev{Local explanations for $\Phi$ variations were first reported in the early 1990s by \cite{1992A&A...264..415T} for the hot spot in 3C~194, and later, e.g., by \cite{2020ApJ...903...36S} and references therein, to explain extreme variations in $\Phi$ across Cygnus~A spanning thousands of \radmm.  Other examples of local Faraday contributions have been suggested for the depolarised patches in Fornax A \citep{2018ApJ...855...41A}, bands of low and high \rev{$\Phi$} in 0206+35, 3C~270, 3C~353 and M84 \citep{LG2}, \rev{and for other systematic Faraday patterns \citep[e.g.,][]{LG1,LG3,2022ApJ...937...45A}}. 
  Recently, \cite{2023arXiv231112363J} have used magnetohydrodynamic (MHD) simulations of active galactic nuclei (AGN) jets to develop \rrev{new} tools for using Faraday variations as probes of their environments. }



\rev{In this paper, we introduce a new visualization technique to allow a more systematic examination of the local Faraday effects in radio galaxies.  The technique itself, introduced in Sec. \ref{sec:tech}, is quite simple, creating Faraday cubes similar to  the $\mathds{F}$ cubes from Faraday synthesis, and then examining them from different angles, either manually or through animations.   We then present a number of \emph{science use cases} in Sec. \ref{sec:use} to first distinguish unrelated foreground from local Faraday media, and then to explore different physical connections between the synchrotron-emitting and magnetized thermal plasmas.  We point out the kinds of science questions these case studies raise for further study, along with a few new science insights. In Sec. \ref{sec:discuss}, we summarize the technical and scientific developments needed to make the technique more objective, robust, and powerful.  Concluding remarks are made in Sec. \ref{sec:concl} . }


\section{Technique}\label{sec:tech}
\rev{The purpose of the technique is to look for correlations between the structures seen in polarized intensity images and the variations in $\Phi$. }
\rev{In many situations, this is very difficult using the pairs of maps that are published, respectively indicating the Faraday depth, $\Phi$(RA,Dec), and amplitude, $P$(RA,Dec), of the local peak in the Faraday spectrum. }  In this paper, we suggest re-creating pseudo-3D cubes $\mathds{P}$(RA, Dec, $\Phi$) from this pair of maps, providing an additional powerful diagnostic of the underlying Faraday structure. We note that the $\mathds{P}$ cubes lose information about complexity that is present in the full Faraday cubes $\mathds{F}$(RA, Dec, $\Phi$), but also gain simplicity when spurious sidelobe structures or faint Faraday components are eliminated. \rev{$\mathds{P}$ cubes can also be constructed where the  $P$, $\Phi$ maps were created from narrow-band multi-frequency observations and no full Faraday cubes are available.}


To create the pseudo-3D cube, $\mathds{P}$, one first chooses the range of Faraday depths $\Phi_1$ to $\Phi_2$ to be displayed, and the number of pixels $n_{\Phi}$ along the $\Phi$ axis. \rev{  Each pixel $k$ on the $\Phi$ axis corresponds to an interval $\pm \frac{\delta\Phi}{2}$ around a specific $\Phi_k$ where $\delta\Phi = \frac{\Phi_2-\Phi_1}{n_{\Phi}}$.  Indicating (RA,Dec) with their pixel coordinates, $(i,j)$, yields} 
    
\rev{$$\mathds{P}(i,j,k) =  \left\{ 
  \begin{array}{ c l }
    P(i,j) & \quad \textrm{if } \Phi(i,j)=\Phi_k \pm \frac{\delta\Phi}{2}\\
    0                 & \quad \textrm{otherwise}
  \end{array}
\right.$$}

\rev{For display purposes, it is useful to smooth along the $k$ axis with some width $\Phi_{sm}$, which can \rev{be used to} provide some indication of the uncertainty in $\Phi$. }It is important to resist the obvious, but incorrect choice of $\Phi_{sm}$ to be equal to the width of the main peak in the Faraday spectrum, since the errors in $\Phi$ are smaller than that by a factor of $2\times$ the signal/noise. Illustrations of the effect of choosing different values for $\Phi_{sm}$ are shown in Appendix A.
\begin{figure}
    \centering
   \includegraphics[width=3.5in]{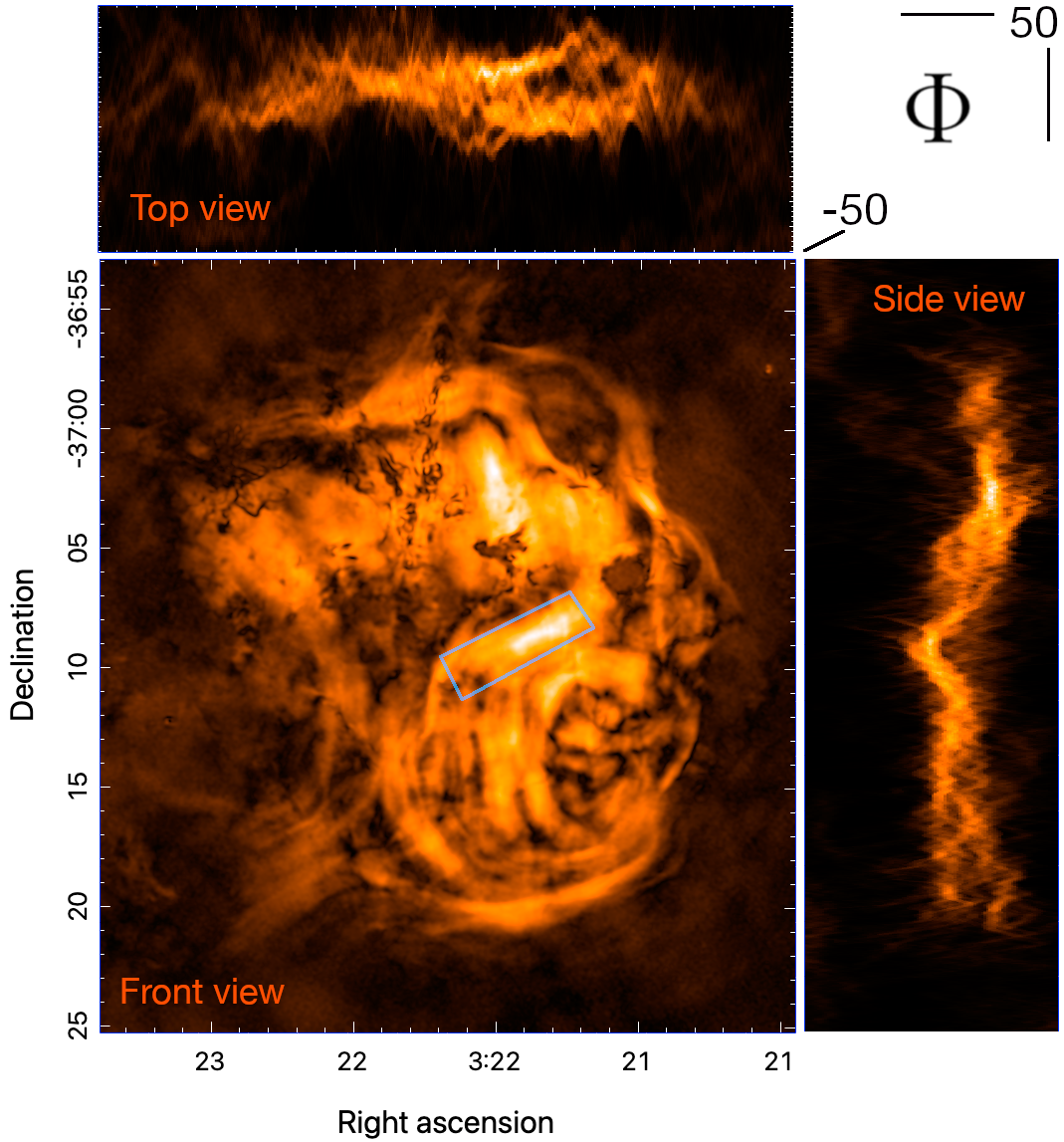}
    \caption{\rev{2D projections of the \PS~ cube of the western lobe of Fornax~A, from the full movie.} Bottom left: polarized intensity of the western lobe of Fornax A, from \citet{2021PASA...38...20A}. The feature enclosed by the cyan rectangle is discussed in Sec. \ref{sec:cross}.  Top:  the ``top view," i.e., the projection of the $\mathds{P}$(RA, Dec, $\Phi$) cube onto the (RA, $\Phi$) plane.  Right: the ''side view," i.e., the projection of the $\mathds{P}$(RA, Dec, $\Phi$) cube onto the (Dec, $\Phi$) plane. In both images the range of $\Phi$ goes from -50 to +50 \radmm. \rev{The smoothing width is 5\radmm. } An animated version rotating through the top and bottom left panels is available in the ancillary files.  The animation is 10 seconds long, and projects the cubes at viewing angles from 0 to 90 degrees around the RA axis. The $\Phi$ axis label is projected into a single position at the beginning of the animation; it becomes visible as the cube rotates. 
    }

    \label{fig:fts}
\end{figure}

Figure \ref{fig:fts} shows the western lobe of Fornax~A. \rev{Examination of the accompanying movie which views the cubes from different angles, shows the existence of long, coherent structures in (RA,Dec,$\Phi$) space. Such features are important for understanding the MHD behaviors of jet-inflated lobes. These structures  can also be seen, partially overlapping each other, in the two projections in Fig. \ref{fig:fts}.}   At the top, we see the view of \PS~ from the top, i.e., projected along the Declination axis onto the (RA, $\Phi$) plane. To the right, we see the view from the side, i.e., projected along the RA axis onto the (Dec, $\Phi$), plane. \revv{These two projections are somewhat arbitrary in the sense that they do not necessarily correspond to the structure of the source.  For each source under investigation, movies and interactive changes in view should be used to find the most useful projections. As shown below for our first science use case, Fornax~A, features can then become apparent that are extremely difficult to recognize in other ways.}

 We provide a python tool to create \PS~ cubes from matched pairs of polarized intensity \rev{(P)} and Faraday depth ($\Phi$) maps, \rev{as described in Appendix B}.  Movies, such as presented here, \rev{are an essential part of looking for the signatures of different physical situations, and} can be created \rev{from \PS~} by using $SAOImageDS9$, or other applications.  \rev{In particular, we recommend use of \href{https://github.com/Punzo/SlicerAstro/wiki}{\emph{SlicerAstro}} \citep{2017A&C....19...45P}, a powerful interactive tool for visualizing cubes.  }

\section{\rev{Science Use Cases}}\label{sec:use}
   \begin{figure}
    \centering
    \includegraphics[width=3.5in]{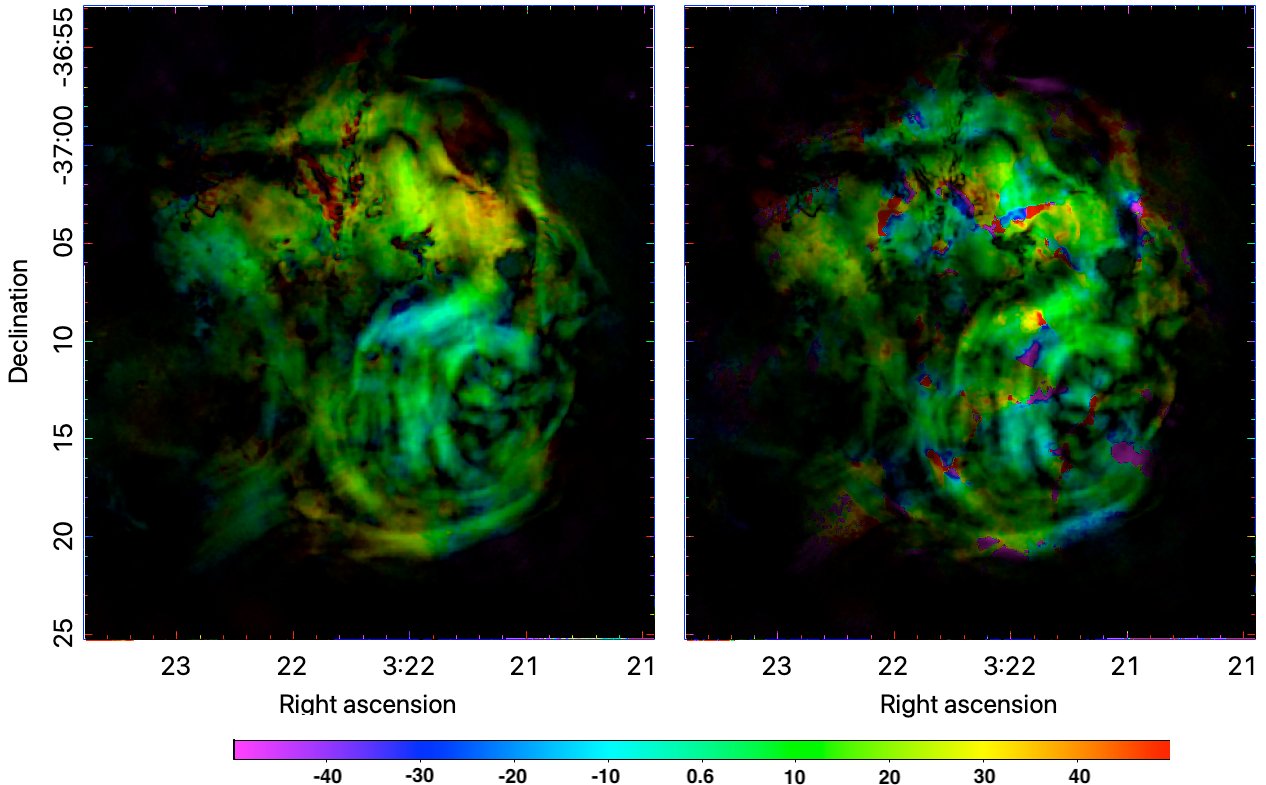}
   
    \caption{The same image of the Fornax A western lobe as in Fig. \ref{fig:ForAW_RMB}, but here with the brightness corresponding to the polarized intensity.  Left: Color-coding is Faraday depth,  the same as Fig. \ref{fig:ForAW_RMB}. Right: Color-coding using  a ``mock" Faraday depth distribution from the Fornax~A \emph{eastern} lobe.}
    \label{fig:For2}
\end{figure}

\rev{We now discuss a number of science use cases to illustrate the type of information that can be extracted from \PS. The scientific scope of this paper is limited to illustrating the types of information that can be extracted from \PS, and why those are scientifically important.} 

\rev{The information derived from \PS~ is based on correlations between the polarized intensity structures and the variations in $\Phi$. If no significant correlations exist, then it is likely that the $\Phi$ variations arise in an unrelated foreground screen, and there is no further information that can be derived from \PS. }

\rev{A hint of the (P, $\Phi$) correlations in Fornax~A can be seen in the left panel of Fig. \ref{fig:For2}, using a different display of the same $\Phi$ map shown in Fig. \ref{fig:ForAW_RMB}. Here,  the brightness in the image indicates the brightness of the polarized intensity map \rev{P} at each position.  }

In the right \rev{panel of Fig. \ref{fig:For2}} is the same polarized intensity image, but this time color-coded by the Faraday depth in an \emph{unrelated}  screen (designated here as ``mock").  The mock screen is formed from the \emph{eastern} lobe of Fornax~A, spatially scaled by a small factor (13\%), to cover the western lobe.  This ``mock" Faraday screen thus has, to first order, the same distribution of Faraday depths and the same spatial scales as the true screen.\footnote{The similarity between the Faraday depth distributions of the two lobes is not necessarily true for all sources, so this type of experiment must be used judiciously.}  \rev{Most of the} correspondence between polarized intensity and Faraday depth structure seems to have disappeared, which it should \rev{if the two are not related.  However, given that there are significant large scale variations across the $\Phi$ map, some accidental correspondences can and do appear.}  

The effects seen in Fig. \ref{fig:For2} are subtle, \rev{however, and we propose the use of \PS~  to better identify relationships between P and $\Phi$.  Our first science use case includes distinguishing between an unrelated intervening Faraday medium and a local one mixed with the synchrotron emitting plasma.} 

    \subsection{\rev{Case 1: Mixed thermal and synchrotron plasmas}} 
    \label{unrelated}


 \rev{\ul{Science context.} There is a substantial literature on filamentary structure in radio lobes. \cite{1984ApJ...282L..55V}  detected fine-scale structures in 3C~310, with high fractional polarizations.  They suggested that these represented ``bubble" boundaries within the diffuse lobes.   \cite{ 1984ApJ...285L..35P} revealed a "wealth" of filamentary structures within the wide lobes of Cygnus~A.  Other suggestions of filamentary structures include the Seyfert-starburst galaxy NGC~3079 \citep{2019ApJ...883..189S} and NGC~6068 \citep{2023A&A...677A...4C}.   \cite{A194} found evidence for mixed synchrotron filaments embedded in a thermal plasma in the southern lobe of 3C~40B.} 

\rev{Here, we examine Fornax~A, whose lobes have unprecedented filamentary structures \citep{1989ApJ...346L..17F}, which are polarized with lengths spanning most of the lobe.   Correlations between depolarization and total intensity structures are presented in \cite{2018ApJ...855...41A}. Examination of \PS~  gives us the opportunity to disentangle the 3D structures within these interspersed and interacting thermal and relativistic plasmas.}\\

\begin{figure}
    \centering
    \includegraphics[width=3in]{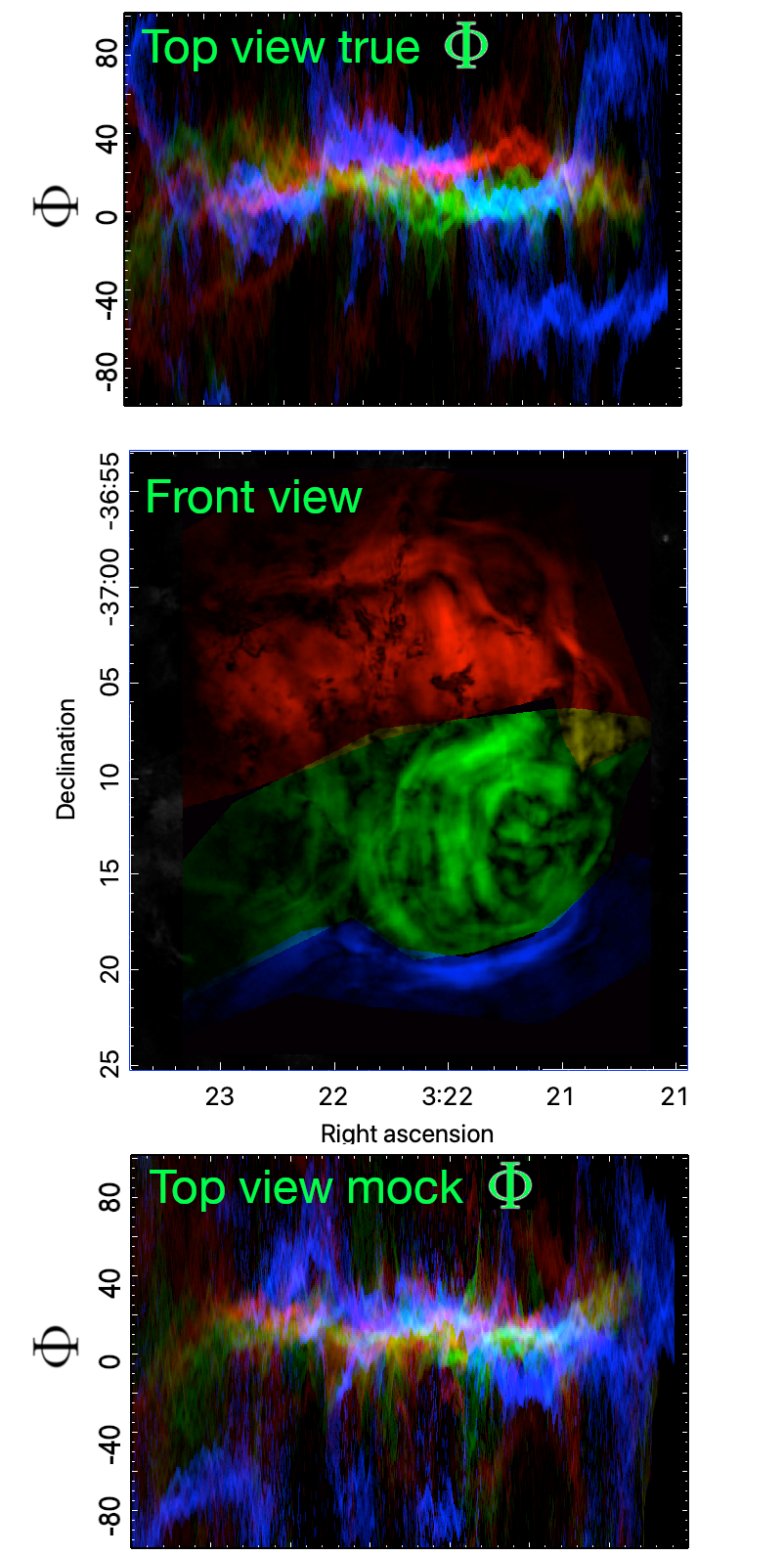}
    \caption{Middle: polarized intensity of the western lobe of Fornax A, with colored bands to distinguish emission from different regions.  Top:  the "top view" showing the projection onto the (RA, $\Phi$) plane, using the true Faraday distribution, showing an expanded  $\Phi$ range from the top panel in Fig. \ref{fig:fts}, with the addition of the color information. \rev{The smoothing width is 5\radmm. } Bottom: the "top view" projection along the (RA,$\Phi$) plane, but using the ``mock" Faraday distribution.}
    \label{fig:dd2}
\end{figure}

 \rev{We use two projections from \PS~  to illustrate the difference in appearance between local and unrelated foreground screens, using the actual and mock $\Phi$ maps described earlier.} Figure \ref{fig:dd2} presents the \rev{same} face-on polarized intensity map \rev{of  Fornax~A West as in Fig. \ref{fig:fts},} along with two versions of the top view in $(RA, \Phi )$ space.  To allow the structures arising from different regions to be seen more clearly, we color-coded the polarized intensity into three different bands.  The color-coding is the same in the two  ``top views".   In the upper "top view", \rev{from the actual or ``true" $\Phi$ maps,} we see the same structures that are visible in the Fig. \ref{fig:fts} "top view", but here we can see that each color band has its own long coherent features \rev{which are even clearer in the movies.}

    In the lower ``top view" projection, the Faraday structure is the ``mock" one.  In most places the structures are heavily overlapping and not cleanly separated as in the true Faraday case.    This arises because the mock Faraday variations are not correlated with individual polarized intensity structures in the West lobe.  Occasional features with larger scales are seen, and  are expected even at random, \rev{when} the Faraday variations have large angular scales.  

     \rev{As discussed in more detail below, foregrounds from the Galaxy can contribute to the observed Faraday variations.  Since they are unrelated foregrounds, they will not be correlated with the structures in P; when they are large enough, however, they could mask underlying correlations.  Fornax~A is large in angular size, but given its high latitude and anti-Galactic direction ($l=240.2^{\circ}$, $b=-56.7^{\circ}$), 
coupled with its location} to the south-west of and outside of the Orion-Eridanus superbubble, we expect a \rev{small or} negligible Galactic contribution to the observed variations in $\Phi$ ($\lessapprox5$\radmm~  on the scale of the lobes and below, based on an RM structure function analysis in the Fornax A region by \citealp{Anderson2015}).  

    \rev{After consideration of possible Galactic effects, the critical test is the comparison of the true and mock views. The separation of different structures visible in the true "top" ($\Phi. RA)$ view, but not in the corresponding mock view, indicates a significant correlation between P and $\Phi$ structures.  This then} leads us to the conclusion that the filamentary polarized structures in Fornax~A West are embedded in a lobe-filling thermal plasma. \rev{It} confirms and extends the suggestion by \citet{2018ApJ...855...41A} for a local thermal plasma as the cause of the depolarised patches in the lobe. \rev{The long lengths of the filaments can now be disentangled from their overlapping structures  when viewed face on, and the scale sizes of the magnetic field variations in the thermal plasma can now be characterized.}
    
\begin{figure}
    \centering
    \includegraphics[width=2.0in]{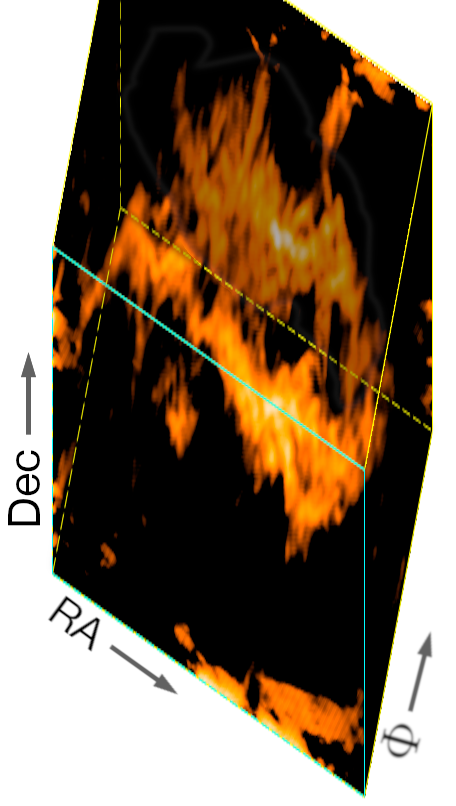}
    \caption{\rev{The \PS~ cube of Fornax A West viewed from an angle of 65$^{\circ}$ above the plane of the sky, and with the major axis rotated towards us by $\sim$20$^{\circ}$ from the plane of the sky.  These angles are in a 3 dimensional space where 1~\radmm = 5 \arcsec. The axes have been stretched to preserve the RA, Dec aspect ratio. }}
    \label{fig:For3D}
\end{figure}

    \revv{Examination of the Fornax~A West cube from all angles led us to a new and important insight. Around one particular angle, the scattered filaments "collapse" into two coherent parallel structures. This is consistent with the bulk of the polarized emission being located in two broad bands along the major axis, as seen in Figure \ref{fig:For3D}, as opposed to being scattered throughout the volume or surface of the lobe.  Most of the northern portion of the lobe, from our normal face-on view,  is at higher (further) Faraday depths, while the bulk of the southern portion is at lower (closer) depths.  }
    
    \revv{The existence of these polarized bands appears clear,  but the scaling, and even sign of the ($\Phi,s$) mapping is unknown; these patterns might therefore be reversed and stretched arbitrarily.  The identification of these polarized emission bands shows the importance of  viewing the \PS~ cube from all angles. In this case, finding an orientation where the emission  ``collapsed'' into compact structures was the key; other patterns may also emerge as more sources are explored.}
    
    \rev{The above findings are based on a relatively uniform magnetic field and electron density within the lobe.  Whether these are physically plausible quantitatively, as well as the implications of banded polarizations are important areas for further study.} 

    \subsection{\rev{Case 2: Local intervening thermal plasmas}} 
    \rev{\ul{Science context.} In Case 1, the thermal plasma and the synchrotron structures within the radio source were mixed on macroscopic scales. A different situation was first discussed by \cite{1962PrCmg...8...48K}, suggesting that NGC~5128 (Cen~A) was interacting with a hypothetical external intergalactic medium. Interactions between outflowing synchrotron plasma and emission line material were later discussed by \cite{1984ApJ...276...79V,1984ApJ...277...82V}, including cases where star-formation was triggered \citep{1985ApJ...293...83V}. Through the identifications of unusual patterns of rotation measures, evidence emerged for interactions of the radio and surrounding plasmas in a number of individual radio galaxies \citep{Carilli1988,1990ApJ...357..373B,LG1,LG2,LG3,RB,2018ApJ...855...41A,2022ApJ...937...45A}. Magnetic draping, \citep[e.g.][and references therein]{2019A&A...622A.209A}, illustrates another type of Case 2 situation, where the thermal material surrounding the radio emitting structure can be studied. More recently, \cite{2023MNRAS.520.4427M} suggested a connection between the bases of the lobes in 3C~34 and 3C~320 and dense regions of the surrounding thermal medium. }
    

 \begin{figure}
    \centering
    \includegraphics[width=3.in]{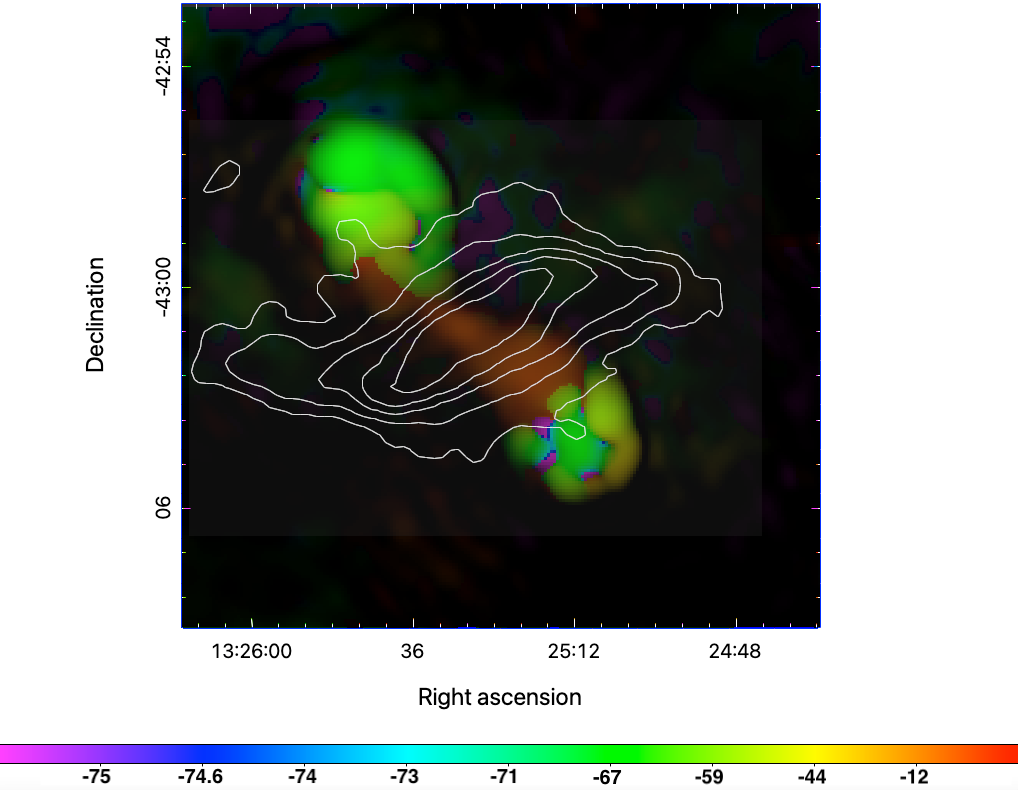}
   
    \caption{Polarized intensity of the inner double of Centaurus~A color-coded by the local value of the Faraday depth.  The contours represent the infrared radiation as observed with Herschel, starting the 3$\times$ the local rms, and each contour level increasing by a factor of 3.} 
    
    \label{fig:cenall}
\end{figure}  

    The inner structure of Centaurus~A was mapped by \cite{CenAinner} showing a $\sim$15~kpc double source connected by jets, roughly perpendicular to the galaxy's dust lane.  Figure \ref{fig:cenall} shows the polarized intensity of this double, color-coded by the Faraday depth, using the ASKAP commissioning data reported in \cite{Centaurus}.  The jets and transition regions into the hotspots have a Faraday depth of $\sim -26$\radmm, while the two hot spots have values near $-61$\radmm, which is the same as the local Galactic foreground \citep{2022A&A...657A..43H}.  Galactic foregrounds are \rev{thus} not likely to significantly affect these results.  
    
\begin{figure}
    \centering
    \includegraphics[width=3.in]{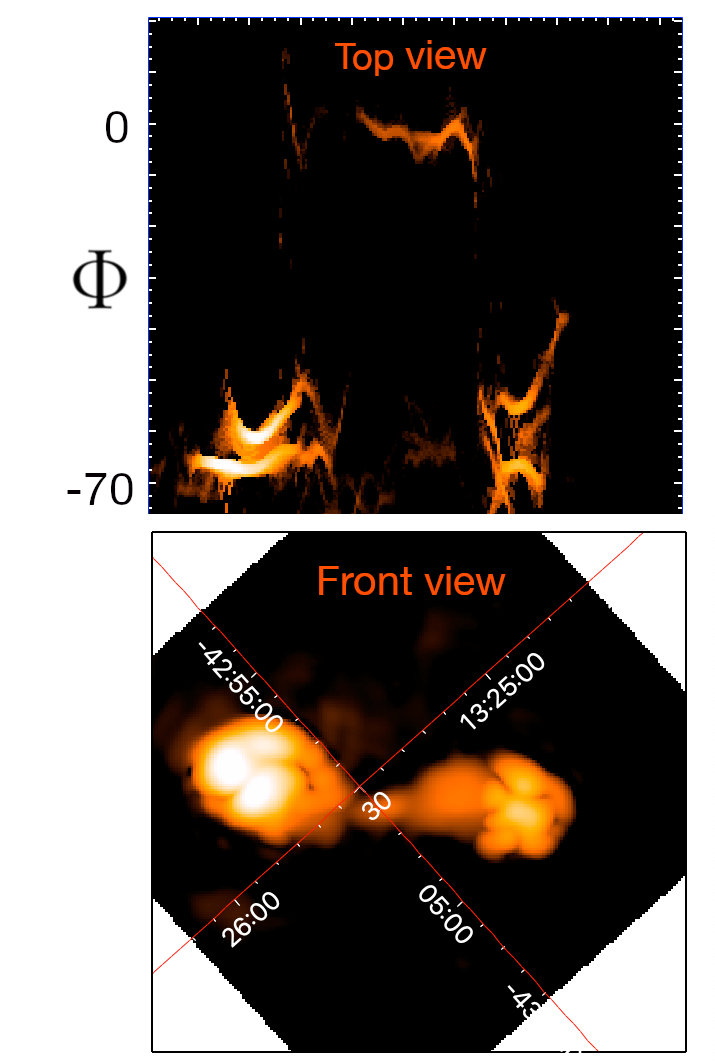}
    \caption{
   Bottom: Polarized intensity of the inner double of Centaurus~A.  Top: The ``top" view in the $\Phi$ vs. position along the major axis, with the spatial scale identical to the bottom panel. \rev{The smoothing width is 2.5\radmm. } An animation rotating through the top and bottom panels is available in the ancillary files.  The animation is 10 seconds long, and projects the cubes at viewing angles from 0 to 90 degrees around the major axis. 
   }
    \label{fig:ftsC}
\end{figure}

    Although the jets and hotspots are continuous in space, the transition between the jet and hotspot Faraday depths is sharp and $\sim10\times$ larger than any variations within the hot spots. \rev{ The distance along the line of sight, $s$, cannot be changing abruptly}, since the jets and hotspots must be physically connected, \rev{Since $\Phi = \int{n_e(\bf{s})~ \bf{B(s)}\cdot d\bf{s}}$, the discontinuity in $\Phi$  must arise in variations of $n_eB$ along the line of sight.}  The discontinuity is also seen dramatically in the top panel of Fig. \ref{fig:ftsC}, showing the Faraday depth $\Phi$ as a function of position along the major axis. As seen in  Fig. \ref{fig:cenall}, Cen~A's dense molecular disk \citep[e.g.,][]{1992ApJ...391..121Q} is the obvious source of the excess Faraday rotation. 

    \rev{In Case 2 situations, the information from the Faraday depth variations can then be used to derive physical properties of the intervening medium, or to check for consistency with existing models. Although detailed modeling is beyond the scope of this paper, we note that the inferred change  in $n_eB$ could arise if the AGN and jets were physically located at the center of} Cen~A's dense molecular disk \citep[e.g.,][]{1992ApJ...391..121Q};  the axisymmetric spiral magnetic field in the disk \citep{2021NatAs...5..604L} \rev{would then provide the necessary}  line of sight component \rev{to produce Faraday rotation}. 
 
\begin{figure}
    \centering
    \includegraphics[width=2.5in]{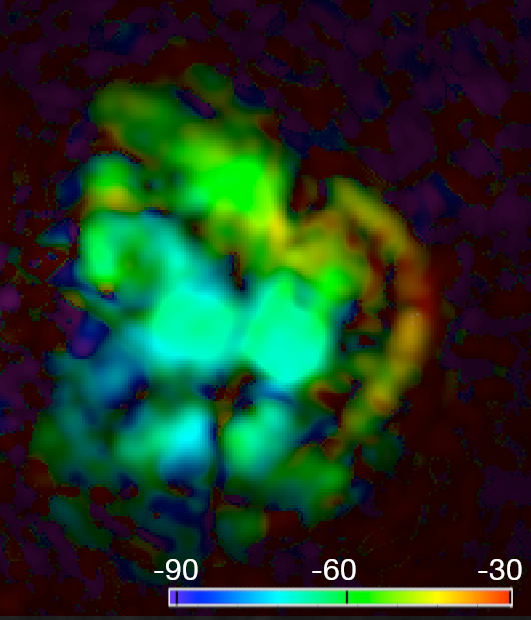}
  

    \caption{High resolution (10") polarized intensity of the southwestern hotspot/lobe of Centaurus~A, color-coded by the Faraday depth, $\Phi$, as shown in the colorbar \rev{in units of \radmm}. }
    \label{fig:ftsCW}
\end{figure}

 In Figure \ref{fig:ftsC}, there is \rev{also} a distinct double structure in $\Phi$ for each of the hotspots.  These double structures arise from the gradients in $\Phi$ along the minor axis, \rev{which can also be seen in the closeup view in Figure \ref{fig:ftsCW}.}  This pattern \rev{is easily} seen in the movie which rotates the viewing angle of the $\mathds{P}$ cube around the major axis. This is consistent with what would be expected from a toroidal component to the \rev{hot spot} fields, with magnetic fields pointing towards us at the top (above the major axis), and away from us below the major axis. It is similar to what is seen in other powerful radio sources with overpressured lobes \citep{2022ApJ...937...45A}.
 
 

 The south-western hotspot also has a thin leading edge, \rev{(see Fig. \ref{fig:ftsCW}}) with a distinct pattern of Faraday structure \rev{which we first spotted in a high-resolution movie}.  The most positive values of $\Phi$ (relative to galactic) are seen at its apex, with values  symmetrically decreasing away from this position.   This  hotspot \rev{was} shown to be  surrounded by a thin shell of X-ray emitting material \citep[Figure 1 in][]{2009MNRAS.395.1999C}\footnote{also see ESO's beautiful  \href{https://www.eso.org/public/images/eso0903a/}{radio/optical/X-ray composite image.}}.   \cite{CenAX} and \cite{2009MNRAS.395.1999C} identify the radio hotspot as the contact discontinuity driving a strong X-ray shock into the ISM. The Faraday patterns at the leading SW edge (Fig. \ref{fig:ftsCW}) \rev{appear} consistent with this, although \rev{detailed modeling is required}.  
   
    \subsection{Case 3:  Revealing 3D structures} 
     \rev{\ul{Science Context.} Determining the 3D structures of radio sources, along with their line of sight orientations, is one of the biggest challenges in understanding their physics. Rotational symmetry along the major axis of straight sources has been built into models, from the earliest \citep{1980Natur.287..208B} to the more recent ones \citep{2018MNRAS.475.2768H}.  On scales of \rrev{$10^{2}-10^{3}$} kpc, the effects of line-of-sight orientation in classical doubles and more recently, for HyMORs sources with terminal hot spots on only one side, have also been of interest \citep{2003PASA...20...50S,2013A&A...557A..75C,2020MNRAS.491..803H,2022JApA...43...97S}.  For more complicated bent sources, 3D models of e.g., precession \citep{2023ApJ...948...25N} or orbits \citep{2004MNRAS.351..101K,2021ApJ...911...56G} are often invoked. One major tool in  orientation studies has been the asymmetry in  depolarization of the near vs. far lobes in a thermal galactic halo, \citep{1988Natur.331..147G,2021MNRAS.508.1371S}; this is complicated by asymmetries in lobe lengths or in the depolarizing medium \citep[e.g.,][]{1989AJ.....98.1232P}}
     
     { A special case of 3D studies is the line-of-sight orientation of jets; on pc scales, these have been studied using circumstellar obscuration and relativistic beaming, including superluminal motions \citep{1984ARA&A..22..319B, 1995PASP..107..803U,2007ApJ...658..232C}.  The orientation of these smaller-scale jets with respect to their large-scale structure provides fundamental information about long-term stability, possible influences of black-hole mergers, and precession of the jet axis. A recent study by  \cite{2023arXiv231202283U}, e.g., finds that misalignments from the largest scale structures are quite common on pc-scales, with timescales ranging from \rrev{$1-10$} Myr.   X-shaped sources are a particularly interesting case, because jet precession provides a possible explanation \citep{2002MNRAS.330..609D}.}
     
    \subsubsection{\rev{Case 3a:} Jet Bending}
\begin{figure}
    \centering
    \includegraphics[width=3.5in]{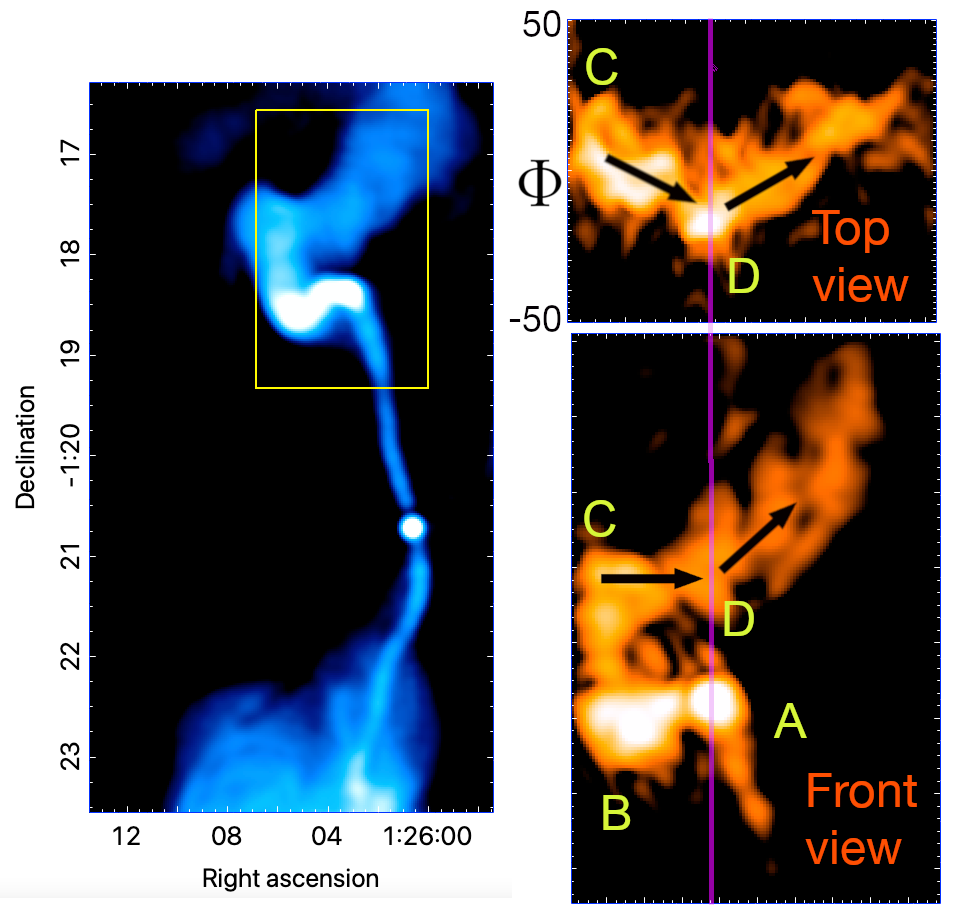}
    \caption{Left: Polarized intensity of the central region of 3C40B, from \citet{A194}. Right: Zoom in on region in yellow box.  Bottom panel is the same front view of polarized intensity. Top panel is the ``top" view of polarized intensity in the \rev{$(\Phi, RA)$} plane, covering the final bend in the northern jet. {At the location of the purple line, the jet changes direction in the plane of the sky (bottom), and the $\Phi$ gradient reverses (top). The smoothing width is 5\radmm. } An animation rotating through  the top and bottom right panels is available in the ancillary files, with viewing angles from face on to $90$ degrees (the top view). } 
    \label{fig:N40}
        
\end{figure}

     \rev{Our first example of identifying  3D structures relies on the comparison between a) the inferred jet flow trajectory in the plane of the sky, with b) the pattern of variations in the accompanying Faraday depths.}  Fig. \ref{fig:N40} shows the inner portion of the radio galaxy 3C40B in the cluster Abell~194 \citep{A194}. \rev{As argued there,} the northern jet in this source goes through three sharp bends, at $A$, $B$, and $C$,  likely the result of an encounter with a dense cloud in the ICM.  After the last sharp bend, the jet expands and slowly changes direction (from W to NW) at the position $D$, marked with the vertical purple line in Fig. \ref{fig:N40}.  As the jet crosses this line, it continues to expand and fade, and the change in direction in the plane of the sky is accompanied by a change in the gradient of Faraday depth.  Before $D$, $\Phi$ \rev{decreases} along the jet;  after this point, $\Phi$ \rev{abruptly increases.}

    \rev{If there were} a dense region in the ICM at location $D$ \rev{causing the jet to bend in the plane of the sky, this would also} cause a \emph{discontinuity} in $\Phi$, which we do not observe.  Instead, $\Phi$ is observed to be continuous, while its \emph{gradient} changes at $D$.  This is consistent with a change in direction of the jet flow along the line of sight, at the same location where a change in jet direction is seen in the plane of the sky.

Because we do not know the local value of $n_eB$, \rev{we cannot directly calculate the scaling between $\Phi$ and} $s$, the distance along the line of sight. \rev{However, we can check the plausibility of this picture} by assuming that the trajectory changes by approximately the same angle along the line of sight that it does in the plane of the sky, $\sim 45\degree$. \rev{Following the prescription of \cite{A194} leads to} a line of sight magnetic field of $1.4 \mu G$, with uncertainties of order unity. This is consistent with expectations of magnetic fields in clusters, so the \rev{changing line-of-sight} scenario is plausible \rev{in this case}.  

\revv{As in the case of Fornax~A, by viewing the cube over the full range of angles, we found that the region of the 3C40B northern jet from B to C ``collapsed" into a thin curvy vertical line when viewed from an azimuthal angle of 43$^{\circ}$ towards the west from the line of sight, in the 3D space where 1\radmm~=~1\arcsec.  This is consistent with the jet flattening into a ribbon-like shape for this portion of its trajectory, perhaps during its deflection by a denser region. The other portions of the jet did not indicate such flattening.}

  \subsubsection{\rev{Case 3b:} Jet orientation} 
\begin{figure}
    \centering
   \includegraphics[width=3.5in]{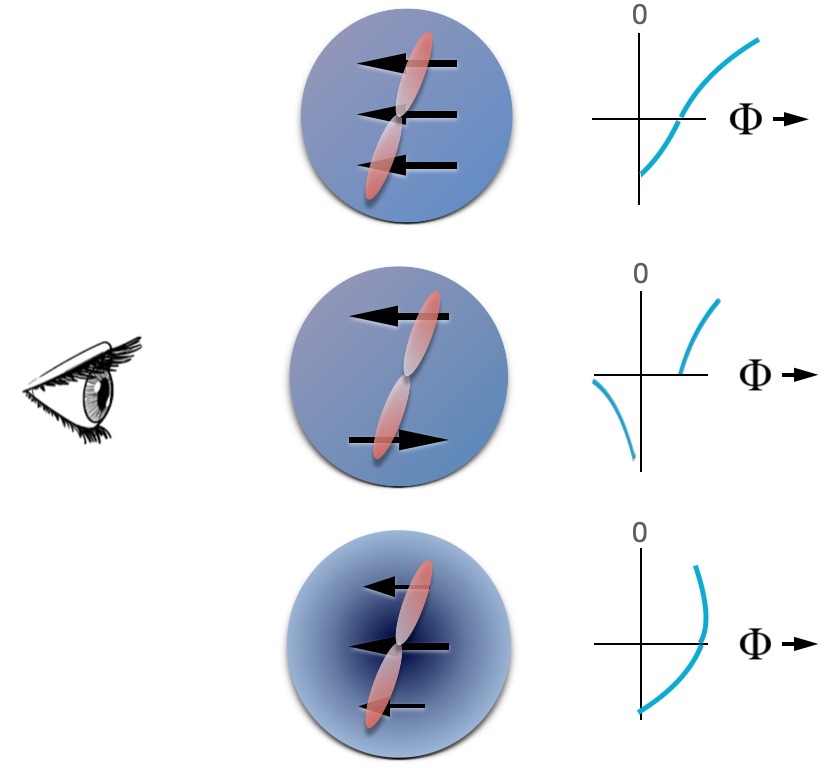}
    \caption{Illustration of Faraday depth along pair of jets for various configurations of thermal density and magnetic field.} 
    \label{fig:RMjet}
  \end{figure}

\rev{Our second example of 3D information involves} the \rev{line of sight} orientation of the jets near \rev{an} AGN, based on their Faraday variations.  Fig. \ref{fig:RMjet} shows three idealized situations for the magnetic and density structures in the \rev{surrounding} galactic thermal medium. \rev{These are not physically accurate, but illustrate the basic observables associated  with different parameters.} We assume in each case that the  Faraday depth \rrev{at the nucleus} is $\Phi_0$ and \ul{that any unrelated foregrounds have been subtracted;} in this case,  the sign of $\Phi_0$ tells us the direction of the magnetic field around the jet.  
  
  In the top example, $n_eB$ is constant throughout the region where the jets are found. \rrev{Although this simplistic case is unlikely, we include it to illustrate the basic expected Faraday patterns. } Because the jet is tilted with respect to the line of sight, \rrev{$\lvert \Phi \rvert$} increases (decreases) with distance from the nucleus for the receding (approaching) jet. 

  The middle example represents the case where \vb is constant in magnitude, but switches direction at the nucleus; \rrev{for simplicity, we show the case where} $\Phi_0 = 0$. \rrev{Such large-scale discontinuities in field directions in a galactic halo have been observed, e.g., in NGC~4631 by \cite{2019A&A...632A..11M}.} \rrev{The key signature of a field reversal is a large discontinuity in $\Phi$ somewhere along the structure. Here, in the special case where the field reversal occurs across the plane of the galaxy,} the receding (approaching)  jet will show an increase (decrease) in $\lvert \Phi \rvert$ with increasing distance from the nucleus. 

  A more interesting and realistic case is shown at the bottom, where $n_eB$ decreases in magnitude with distance from the AGN.  For the approaching jet, \rrev{$\lvert \Phi \rvert$} drops monotonically, because  of both the decreased path length along the line of sight and the lower magnitude of $n_eB$.  For the receding jet, there are two opposing effects; the increased path length with distance from the AGN leads to an increase in \rrev{$\lvert \Phi \rvert$}, while the lower magnitude of $n_eB$ causes it to decrease with distance from the nucleus.  On the receding side, \rrev{$\lvert \Phi \rvert$} therefore rises and reaches a maximum at some distance from the AGN, and then decreases.
  
  We now apply this type of analysis to 
 PKS~2014-55, a 1.6~Mpc long radio galaxy associated with the Seyfert~2 galaxy 2MASX J20180125-5539312, at a redshift of 0.0606. 
 PKS~2014-55 has been previously mapped using MeerKAT at a frequency of 1.3 GHz \citep{PKS2014}, and we use those data in this analysis.
     
     \begin{figure}
    \centering
  \includegraphics[width=3.5in]{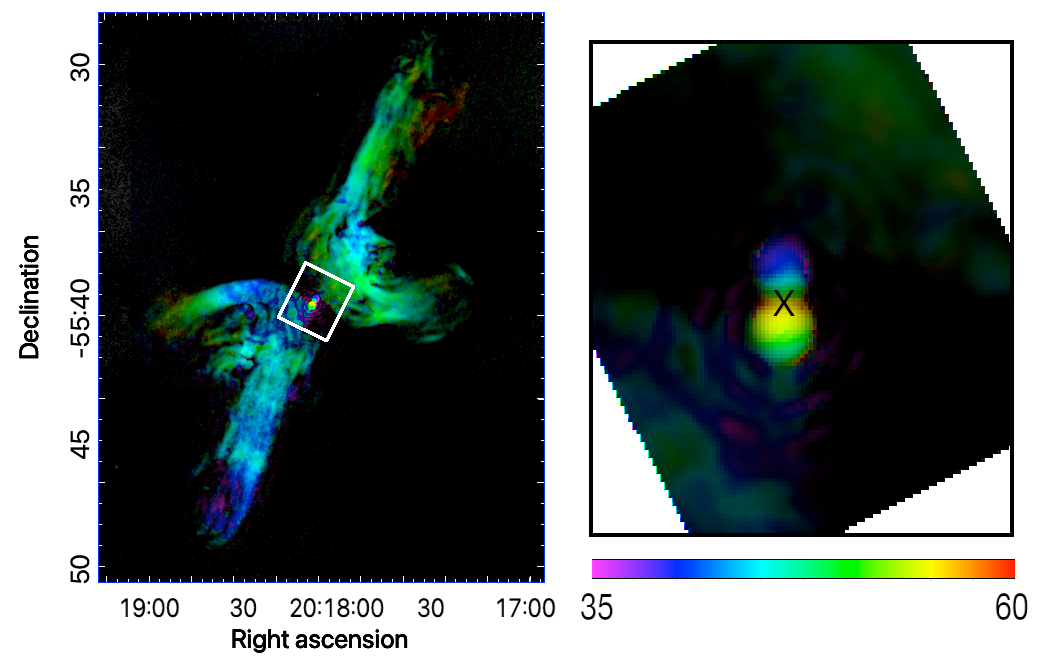}
  \caption{PKS2014-55 in polarized intensity from \citet{PKS2014}, color-coded by Faraday depth, using the same colorbar for both frames. The right frame is rotated and zoomed in on the white box shown on the left. The host galaxy, at 20:18:01.286, -55:39:31.6, is marked with an X.}
    \label{fig:full2018} 
    \end{figure}

 As seen in Figure \ref{fig:full2018}, the Faraday depths \rev{show only small variations on large scales,  with a mean value of} \rev{45\radmm}. The Galactic foreground in this direction, is $\sim+45\pm5$, \rev{so there is little net contribution from the \emph{large-scale}  environment of the radio galaxy.} \rev{However, a} wider range of Faraday depths is seen in the small-scale structures around the location of the host, \rev{while} their mean \rev{is similar to} that of the extended lobes. 
 The \rev{double radio core structure} probes the medium on the scale of $\sim$25~kpc from the AGN, where the ISM and potential halo of the host likely dominate.
  
\begin{figure}
    \centering
         \includegraphics[width=3.in]{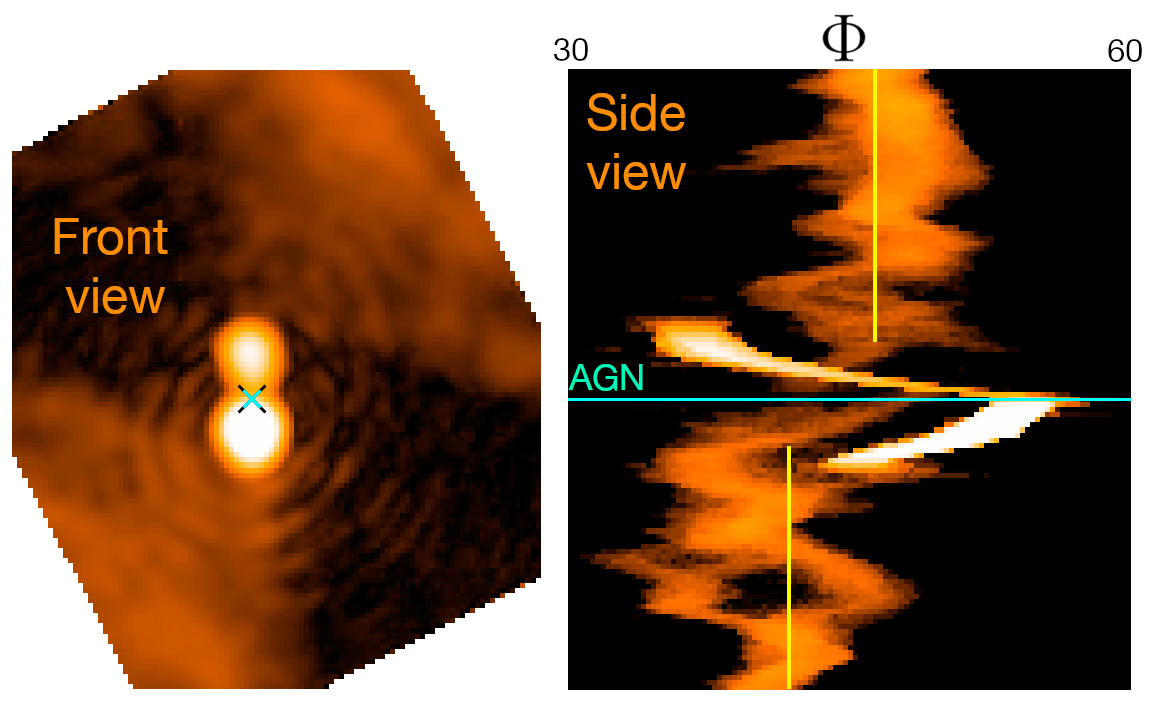}
    \caption{Polarized intensity of the central region of PKS~2014-55 rotated so that the inner double is vertical. Left: Front view, in the plane of the sky.  Right: Side view, showing the variation of $\Phi$ along the major axis. The smoothing width is 2.6\radmm. An animation rotating through the left and right panels is available in the ancillary files.  The animation is 10 seconds long, and projects the cubes at viewing angles from 0 to 360 degrees around the major axis. 
    }
    \label{fig:fts20}
\end{figure}

\begin{figure}
    \centering
   \includegraphics[width=2.7in]{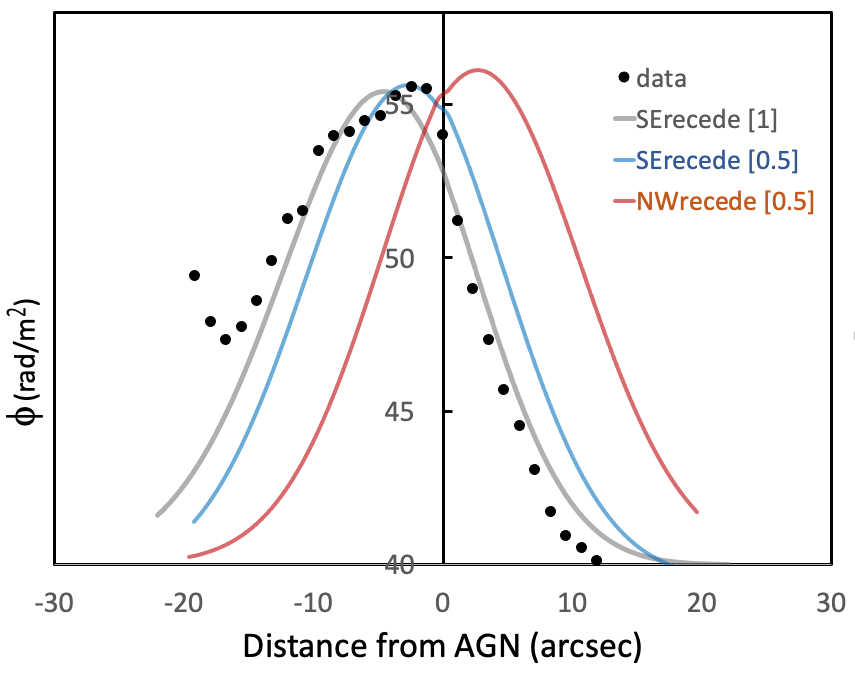}
    \caption{Faraday depth as function of angular distance from the host for PKS~2014-55. Negative numbers along the horizontal axis are to the SE.  The observed behavior is similar to that in the bottom panel of Fig. \ref{fig:RMjet}, with the \rev{peak Faraday depth in the receding, SE,} lobe.   The distance of the peak from the core depends on the ratio of $sin(\theta)$ (the angle of the jets to the line of sight) to the half-power half-width of the Faraday depth \rev{ assumed to be a} 2D Gaussian distribution. Two models \rev{differing by a factor} of two in their half-widths are labeled as [1] and [0.5].  The model also includes an amplitude normalization for $\Phi$, reflecting the unknown magnitude of $n_eB$. } 
    \label{fig:PKS2014plot}
\end{figure}

No total intensity or polarized radio core is seen exactly at the position of the host galaxy, so we assume that the two slightly resolved radio components straddling this position are either jets or an inner double.  The significant Faraday structure in these jets can be seen in the side view of Fig. \ref{fig:fts20} and \rev{in the} accompanying movie.  The NW jet's depth is the same as that of the lobes and \rev{their} ``wings", while the SE jet depth extends to larger  $\Phi$ values. This is consistent with the NW jet emerging in front of the thermal emission in the core, \rev{while} the \rev{Faraday depth} to the SE jet increases away from the core, \rrev{peaking a few arcsec to the south, as seen in Fig. \ref{fig:PKS2014plot}}.  \rev{With the assumptions of a uniform field direction near the host,}  the NW jet would \rev{then} be approaching us, and the SE jet, receding.  

\rev{To model this more quantitatively, we examine} the Faraday depth along the major axis of the double structure (Fig. \ref{fig:PKS2014plot}). The lack of a monotonic trend in $\Phi$ rules out the scenario in the top panel of Fig. \ref{fig:RMjet}. It also rules out the scenario in the middle panel, where the magnetic field switches direction across the core,  because no large jump is observed.  

The data are consistent, however, with the bottom panel, where $n_eB$ is highest at the center, as expected physically, and \rev{where} \vb maintains the same orientation throughout. This leads to the observed offset peaked pattern in $\Phi$ (Fig. \ref{fig:PKS2014plot}); \rev{the plotted} data are accompanied by several models of a tilted jet embedded in a centrally peaked thermal plasma.  It \rev{ confirms the new result that} the NW lobe is approaching us.  \rev{and that the field is pointed towards us. This is also consistent with the expectations from the larger scale emission,  since the average value of $\Phi$ in the \rev{small} double is $+14~$\radmm~ above that of the lobes.} The $\Phi$ pattern would be negative, mirror reflected, if the sign of the magnetic field were reversed.  

\rev{With a more physically realistic model of the thermal medium density and magnetic field, it would be possible to use these results to constrain the relationship between the tilt angle and the radial extent of the medium.  Additional modeling of the large-scale Faraday structure would be useful to confirm the conclusions of \cite{2020MNRAS.495.1271C} that the offset ''wings" are likely due to deflections of the backflow, as opposed to precession of the jets.}

\subsection{Case 4: Jet internal magnetic field structure} 
\rev{\ul{Science context.} \rev{In the above cases, the thermal and synchrotron emitting plasmas were either completely separated or interspersed, but not mixed on microscopic scales. In those cases, the magnetic fields can be quite different in the two plasmas. }  In other situations, thermal plasma can be \emph{fully mixed} with the relativistic plasma, so that the Faraday depth depends on the same magnetic field that gives rise to the polarized synchrotron radiation.   Such mixed plasmas can be difficult to detect, because the ``internal" , broadened Faraday structure results in depolarization that cannot be reduced using narrower bandwidths or higher angular resolutions.\footnote{ \rev{For broadened Faraday spectra, the parameter $W_{max}$ indicates when the sensitivity drops by a factor of two} \citep{RudnickCotton}. \rev{$W_{max}$} ranges from $\sim$150\radmm~ for VLASS \citep{Lacy2020} to only 0.5\radmm~ for the LOFAR HBA \citep{vanH2013}.  $W_{max}$ is much smaller than the commonly used values of ``Faraday width"  \citep{Brentjens2005}. }} 

\rev{When jets can be resolved transversely, their 2D  magnetic field configurations can be studied.} At pc scales, e.g., the MOJAVE survey \citep{Hovatta_2012} found four examples of transverse \rev{$\Phi$} structure, providing early evidence for the presence of toroidal or helical fields; such geometries are now well-established \citep{2021Galax...9...58G,2022ApJ...924..122G}. \rrev{However, simulations of such gradients by \cite{2012AJ....144..105H} show that spurious features can result, but that they essentially disappear when the jet is at least two beams wide, and the signal:noise is $>3\sigma$ along the observed gradient.}

Spine-sheath structures can also \rev{be observed} on these scales \citep{2022JApA...43...97S}. Moving to kpc scales, there is sufficient resolution in some cases to resolve the \rev{projected, transverse} structure of their magnetic fields. \cite{2013MNRAS.432.1114L} describe the observed magnetic field geometries and their connection with the inferred velocities as jets undergo regions of expansion. In NGC~315, for example, \cite{10.1093/mnras/stt2138} \rev{find that the fields} evolve into a toroidal structure along the jet edges with a \rev{central}  poloidal component.

  This combination of poloidal and toroidal components is also reflected in the helical magnetic field configuration in M87 \citep{M87P}. Here, the \emph{emissivity} also shows the double helical pattern, perhaps sustained by Kelvin-Helmholtz instabilities. \rev{They also used Faraday rotation variations across the jet to illuminate the fields' line of sight components, supporting} the presence of the toroidal component.  In this paper, \rev{we probe} additional details of M87's magnetic field structure \rev{using} the \PS~ analysis.  
\vspace{ -0.1in}

\subsubsection{Magnetic field structure in the M87 jet}
\begin{figure}
    \centering
   \includegraphics[width=3.in]{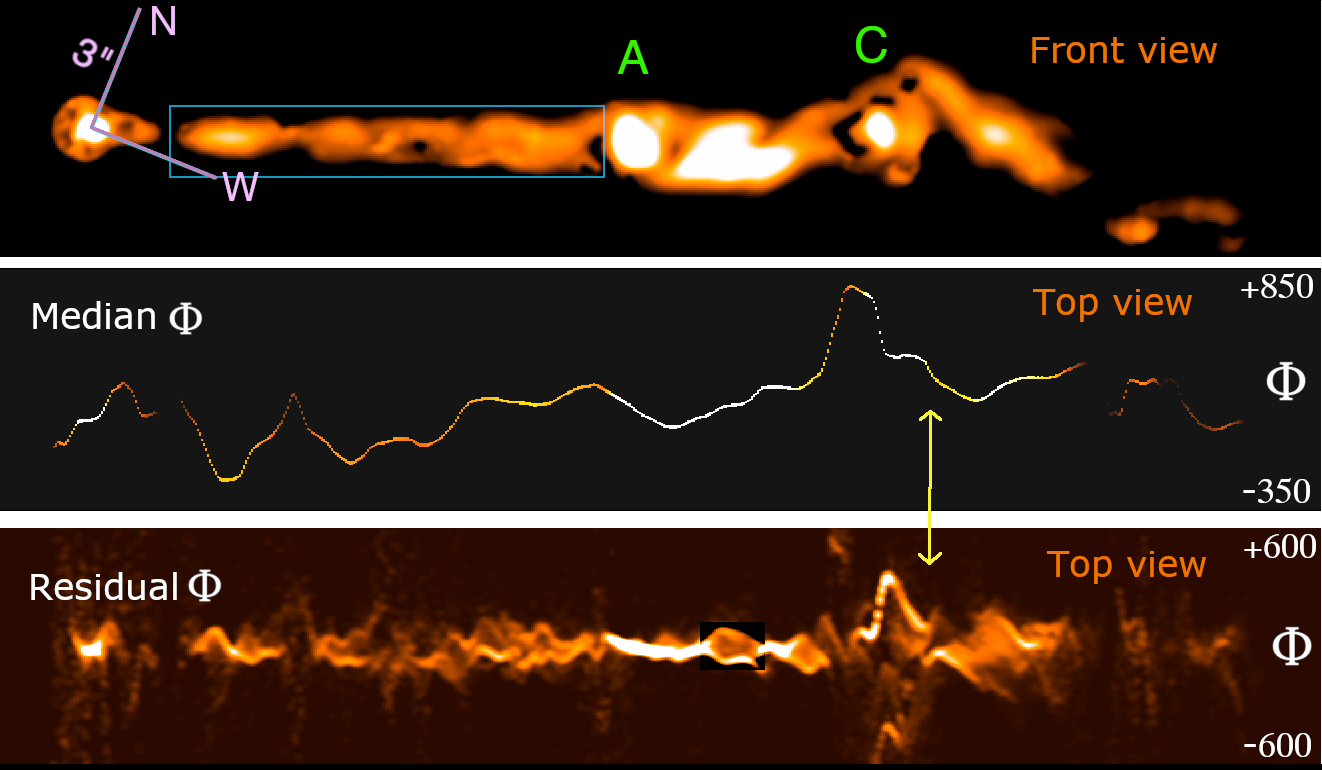}
    \caption{Polarized \rev{emission} of the M87 jet, \rev{from \citet{M87P}}, using the 18GHz bandwidth data from C to Ku Bands, 0.43" resolution.  Top: \rev{Polarized intensity,} in the plane of the sky, rotated by -22$^\circ$.  Middle: Median value of  $\Phi$ across the jet at each location along the jet.  The brightness corresponds to the local polarized intensity. Bottom: Top view of the \rev{\PS~ } cube, \rev{constructed after removing the local median $\Phi$ value}. \rev{The smoothing width is 25\radmm. } The cyan box shows the location of the ``conical" jet region discussed further below.  The yellow arrows indicate a discontinuity in the median and residual Faraday depths.  An animated version rotating through the top and the bottom panel is available in ancillary files.  The animation is 10 seconds long, and projects the cubes at viewing angles from 0 to 90 degrees around the major axis.} 

    \label{fig:M874frame}
\end{figure}

\rev{The polarized and Faraday structures of} M87 (Fig. \ref{fig:M874frame}), are based on the high frequency and high resolution analysis of \cite{M87P}.  The median value of $\Phi$ across the jet at each location \rev{shows smooth} variations along the jet, \rev{covering} a large range ($\lessapprox1000$\radmm, see 2nd panel in Fig. \ref{fig:M874frame}).  The largest excursion is in the region around  bright Knot C, where the jet also undergoes a sudden bend. This correspondence between the bend and the large change in Faraday depth makes it likely that the variations are local to M87, as opposed to an \rev{unrelated} foreground screen. Within the area covered by the M87 jet ($\sim 3\arcmin $ at $ l = 284^{\circ}$ and $ b= 74.5^{\circ}$), the Galactic contribution to the observed $\Phi$ variations is negligible, as the region is at high latitude and no star-formation activity is present that \rev{could} cause noticeable variations. 

\cite{M87P} observed significant depolarization even at the relatively high frequencies of 4~GHz\footnote{using the 18~GHz bandwidth from C to Ku VLA bands}.  \rev{This likely results from the large Faraday widths within each beam,} of  $\sim$ 300 rad/m$^2$ in the "conical" region of the jet \rev{near the core} and  $\sim10^3$\radmm~ \rev{further out}. \rev{A thermal plasma mixed  with the synchrotron emitting jet would cause such high values.}    

\rev{In order to explore the mixed thermal/synchrotron plasma, and} the jet's internal magnetic field structure, we remove the mean variations along the jet, and look at the \emph{residual} $\Phi$, i.e., the variations in $\Phi$ \emph{across} the jet at each location.
\cite{M87P} used these $\Phi$ gradients to infer a double helical magnetic field structure in M87, extending out to $\sim1$~kpc from the nucleus.  The third panel in Fig. \ref{fig:M874frame} is the ``top view" of the \rev{median-$\Phi$-subtracted} Faraday cube;  in most locations there is a significant range of $\Phi$ values \rev{across the jet}.  \rev{This is consistent with the expectations for a toroidal (helical) field component.}

\subsubsection{Faraday cross-sections of jets }\label{sec:cross}

\begin{figure}
    \centering
   \includegraphics[width=3.in]{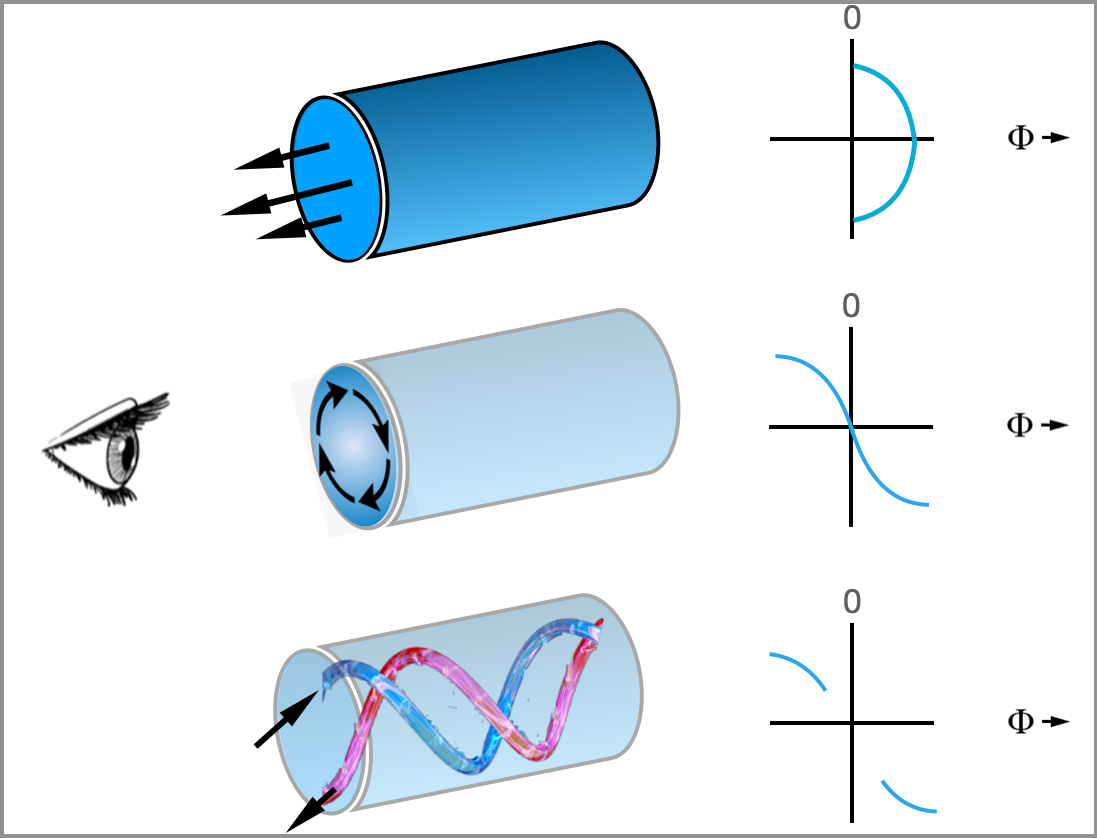}
    \caption{Cartoon showing the ''characteristic" expected behaviors of $\Phi$ as a function of the distance along the projected minor axis of a jet or filament, for three simple magnetic field geometries. The  $\Phi$  axis is with respect to the local \rrev{fore}ground Faraday depth. Top: poloidal; Middle: toroidal; Bottom: toroidal, but confined to a double-helical structure. \rev{In either the toroidal or double-helical configurations, the gradient in $\Phi$ may be offset by an unsubtracted mean background.}}
    \label{fig:RMinjet}
\end{figure}

\rev{To probe the} magnetic structure \emph{within} a jet 
we need a different visualization scheme.  It is useful to focus on the ``side view" ($Y,\Phi$) looking down  the major axis of the jet, where  $Y$ is the position transverse to the jet axis.   Fig. \ref{fig:RMinjet} illustrates the expected results for ($Y,\Phi$) for some simple internal magnetic field geometries. 

\rev{As we turn to a more detailed structural analysis of a jet, it is important to recognize that none of the models presented in this paper are unique.  There is no procedure which maps a set of observations onto a single physical model.  Cartoon models are important, however, to check for consistency with the observations.}

\rev{There are some important further caveats.  In the discussion below, the patterns shown for these ideal cases represent the overall expected trends; the $\Phi$ values are characteristic of the \emph{average} $\Phi$ along each line of sight, not necessarily its value at the peak polarized intensity.  That value depends on the actual emissivity profile of the jet as a function of radius and the detailed magnetic field geometry.  The cartoons represent a starting point for more detailed modeling.} 

\rev{Recognizing these caveats, we} first consider the contribution from a poloidal field at a slight angle from the plane of the sky.  
For a uniform field strength, \rev{$B$}, and constant electron density, \rev{$n_e$} across the \rev{jet}, the (absolute) Faraday depth will be largest along the axis, and decrease to the edges because of the shorter path lengths \rev{(Fig. \ref{fig:RMinjet}, top panel)}.  The exact shape depends on the behavior of $B(r)$ and $n(r)$, where $r$ is the distance from the axis. 

  In the presence of toroidal fields, (bottom two panels of Fig. \ref{fig:RMinjet}), there is a characteristic quasi-linear trend in $\Phi(Y)$, whose exact shape again depends  on $B(r)$ and $n(r)$. The linear trend with radius reflects the fact that toroidal fields have a relatively stronger line-of-sight component the further one is from the axis.  The Faraday depth tends to level off at the extremes as the integrated path length through the cylinder becomes smaller.  If either the electron density or magnetic field strength strengthen off-axis, as in jet models with sheaths experiencing shear \citep{2014MNRAS.437.3405L,2023MNRAS.519.1872W}, \rev{these patterns would change}. \rev{Note that these models simply provide consistency checks, not a unique determination of the field structure. }


 \begin{figure}
    \centering
   \includegraphics[width=3.4in]{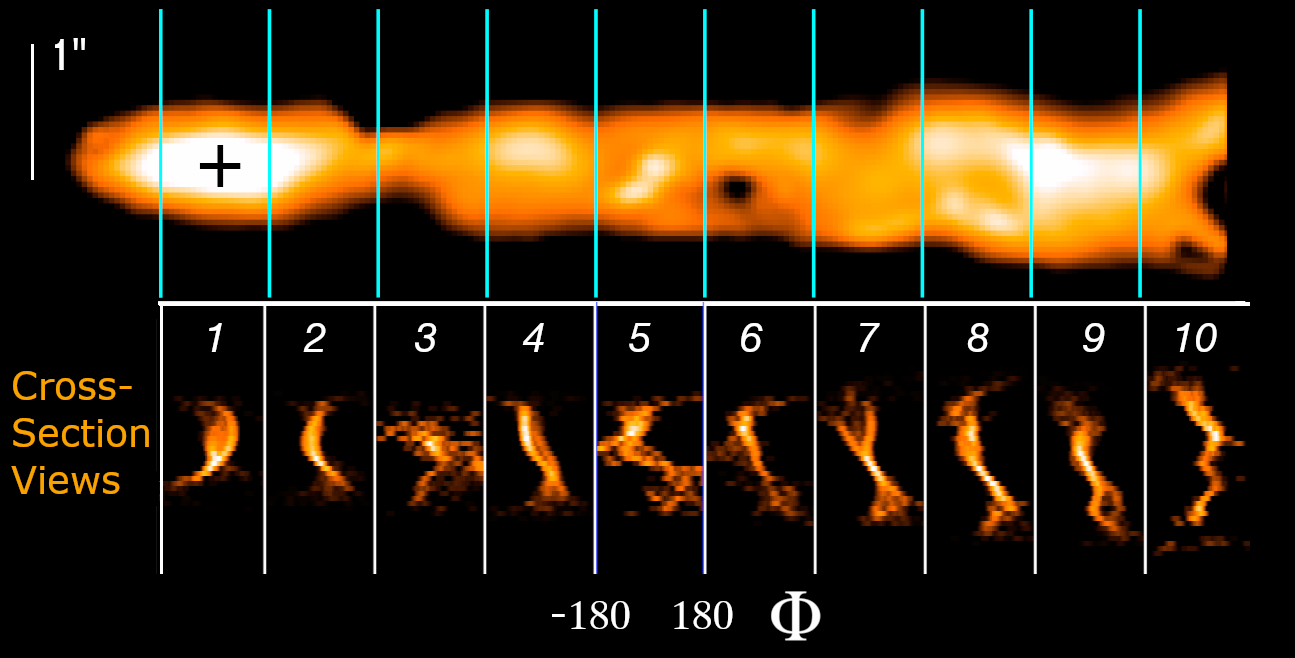}
    \caption{Section of M87 jet 9.5" long (along jet axis) by 2.5" high, at a resolution of 0.45",  covering the region shown in the cyan box in Figure \ref{fig:M874frame}.  Top: polarized intensity.  Bottom: a series of ten cross sectional frames showing the polarized intensity at each value of $\Phi$, with the vertical axis the same as the top panel.  The Faraday structural patterns are to be compared with those in Figure \ref{fig:RMinjet}. \rrev{Only regions 6-10 meet the reliability criteria of \citet{2012AJ....144..105H} - see text.}  \rev{The smoothing width is 25\radmm. } }
    \label{fig:M87slices}
\end{figure}
 \rev{Because M87's jet changes structure along its length, on scales $\geq$10\arcsec, we split up our visualizations to look at a series of cross sections, each 9.5\arcsec long, and}  displayed individually in ($Y$,$\Phi$) space   \rev{(Figure \ref{fig:M87slices}).}  \rrev{Only regions 6-10 meet the full reliability criteria of \cite{2012AJ....144..105H} for transverse gradients. Although the signal:noise is high in regions 1-5, they are less than two beams across. Our discussion of results for regions 1-5 should be considered as suggestive, but not robust.}
 
 \rev{Cross-sections 1 and 2} correspond to \cite{M87P}'s Knot D. Although the jet is only slightly resolved transversely, it  shows significant changes in $\Phi$ across the jet at these locations. The Faraday patterns \rev{here} are close to the ``C-shaped" poloidal example shown in Figure \ref{fig:RMinjet}, \rev{and} appears consistent with the face-on polarization data of \cite{M87P}. The jet's major axis is at a position angle of $-68\degree$, while the magnetic field angle, using the data from \cite{M87P} is $-62.5\degree$; the component of the field, then, projected onto the sky, is consistent with being poloidal, i.e., along the jet.  This is the first evidence, to our knowledge, of 3D evidence for a jet poloidal field. 

Cross-sections ([4], 6, 8, 9) show the clear tilted linear structure of toroidal (helical)  fields. \rev{This is consistent with} the \cite{M87P} double-helix structure, \rev{which we now detect} in the third ($\Phi$) dimension.  \rev{In addition, \cite{M87P} found that the synchrotron emission itself showed the double-helix structure, along with the expected alignment of the magnetic field with the helices, in the plane of the sky.}

\begin{figure}
    \centering
   \includegraphics[width=3.in]{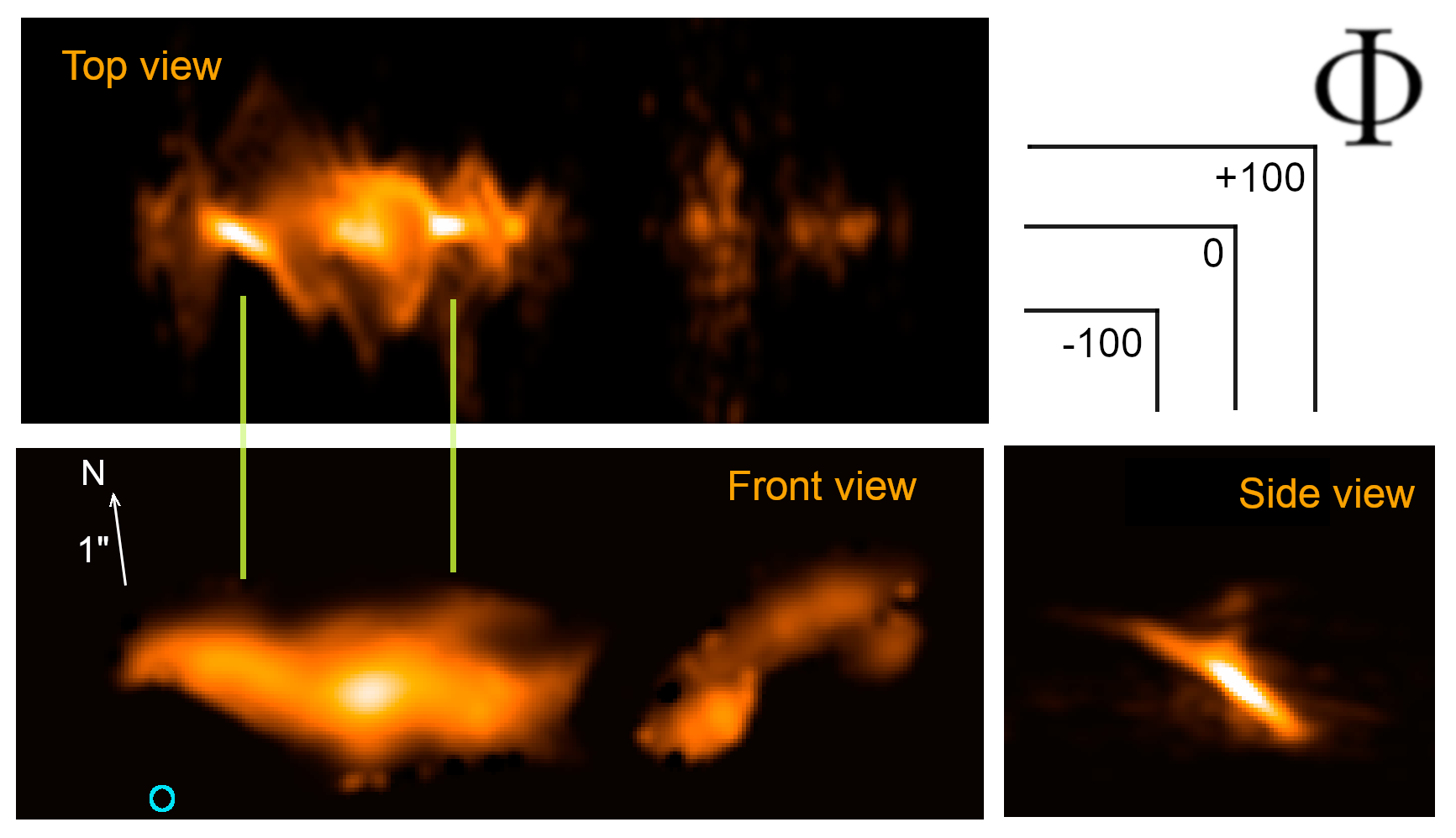}
    \caption{Polarized intensity of M87's far tail, approximately 9" long, with the location of bright Knot C marked with a cyan circle. Bottom left: View from the front.   Top:  the "top view" showing the Faraday depth as a function of position along the jet (matched to bottom left axis along the jet.  Bottom right: ''Side view" looking down the axis of the jet, position transverse to the jet, matched to the bottom left frame. \rev{The smoothing width is 25\radmm. } }
    \label{fig:M87afterC}
\end{figure}

 We turn briefly to  M87's far jet, the  $\sim9$\arcsec region past Knot C (Fig. \ref{fig:M87afterC}). Here, the jet goes through a sharp bend \rev{\citep{1990ApJ...362..449O}}, and there are large excursions in the Faraday depth (Fig. \ref{fig:M874frame}). The ``top view" in Fig. \ref{fig:M87afterC} shows the \emph{residual} $\Phi$ along the jet, i.e., after removal of the \rev{median} variations.  In most locations, the jet has a finite Faraday width, indicating the presence of variations \emph{across} the jet.  In the ``side view," we find that the emission is dominated by a clear, monotonic gradient in $\Phi$ across the jet, the signature of a toroidal/helical field, as shown in Fig. \ref{fig:RMinjet}.  There are also irregular enhancements in brightness along the upper and lower edges in the top view of the far jet, \rev{as would occur for a double helical structure}.  \rev{This is a new finding, since the earlier work \citep{M87P} detected helical structures only in the conical region of the jet, nearer the AGN.}

 \rev{Other regions,} as indicated by the vertical yellow lines in Fig. \ref{fig:M87afterC}, show little variation in $\Phi$ across the jet. \rev{All of the above initial findings suggest useful further work, e.g., considering additional jet components, } the effects of resolution, and smearing in Faraday space due to finite signal:noise, \rev{as well as physical modeling of the evolution of jets with double-helix structures.}



\section{Discussion}\label{sec:discuss}  

  With the rapid growth in Faraday mapping from SKA precursors, and the SKA itself, the number of well-resolved, sufficiently sensitive maps will expand dramatically, \rev{and thus the opportunity to exploit more powerful diagnostic techniques than have been previously available. } 

\rev{We have demonstrated that a wide range of science can be studied using \PS, and that  it can test or constrain simple idealized models for the structure and interactions of the thermal plasma that is adjacent to or mixed with the synchrotron plasma. } Full Faraday synthesis cubes, $\mathds{F}$(RA, Dec, $\Phi$), \rev{can also be viewed in 3D, and offer some of the same diagnostic power \citep{A194}, although they may come at the expense of increased complexity.}   ~ \PS~ \rev{also offers an advantage over $\mathds{F}$ } in terms of resolving Faraday space variations; for full $\mathds{F}$ cubes, the resolution  is set by the width of the Faraday restoring beam \citep[see][]{RudnickCotton}, \rev{whereas for} \PS,  the resolution can be chosen to more closely reflect the accuracy of $\Phi$ in each pixel; see Appendix A. Current and future polarization mapping programs, such as POSSUM \citep{Gaensler2010}, VLASS \citep{Lacy2020}, MIGHTEE-POL \citep{10.1093/mnras/stae169} and Apertif \citep{refId0} should routinely make use of \PS.

\rev{ In addition, \PS~ can be used to exploit the} extensive information on Faraday variations already in the literature, where only  2D images are available.  We \rev{thus} recommend that \PS~ analyses be applied to all existing and accessible pairs of polarized intensity and rotation measure maps, wherever there are a large number of independent spatial beams in both maps, and where the \rev{$\Phi$} variations are clearly detectable above the scatter due to noise.

\rev{Beyond the specific science use cases presented here, the most important aspect of \PS~ is its use as an \emph{exploratory} tool.  As we initially experimented with this new technique, we did \emph{not} target the specific science objectives in the use cases,  such as filaments or jet orientation.  Instead, the science use cases are what \emph{emerged} from our unguided examination of the movies, and it is likely that broad application of \PS~ will uncover other, important science questions.}


\rev{The major challenge in using \PS~ is that it does not provide objective or robust measurements of physical parameters, nor a quantitative way to distinguish between unrelated foregrounds and local effects.   In the remainder of this section, we use what we've learned so far to suggest the most useful areas for further development.}
\vspace{-0.2in}
\rev{\subsection {Algorithm development}}\label{alo}
\rev{The most basic need is for an algorithm to objectively distinguish between unrelated Faraday foregrounds and those associated with the emitting source.  Unrelated foregrounds are present for all sources, from the Milky Way at the very least.  In some science use cases, one must remove the mean foreground Faraday depth, for which all-sky \rev{$\Phi$} maps such as those of \cite{2022A&A...657A..43H} can be used.\footnote {CIRADA (the Canadian Initiative for Radio Astronomy Data Analysis) provides a cutout server \href{http://cutouts.cirada.ca/rmcutout/}{http://cutouts.cirada.ca/rmcutout/} for this purpose.}   The number of background sources suitable for this purpose will be dramatically enhanced with the POSSUM Survey RM Grid \citep{vanderwoude2024prototype}.}

    
\rev{More detailed information on the \emph{local} thermal plasmas comes from studying the \emph{variations} in Faraday depth across the source. To study these local effects, the}   Faraday depth variations in the foreground  \emph{on the scale of the source structures} \rev{must be} small with respect to the local variations.  Fortunately, the Faraday scatter \rev{from the Milky Way} on $\sim$arcminute scales can be quite small.  In the POSSUM Pilot I fields centered at $b=341.7^{\circ}, l=-44^{\circ}$, \cite{vanderwoude2024prototype} found a median \rev{$\Phi$} difference of only $\sim3$\radmm\ between sources with polarized fractions above 3\% up to separations of 15\arcmin. Most radio galaxies are much smaller than this, although nearby sources with a great deal of Faraday structure, such as Fornax~A studied here, are more likely to have significant Galactic contributions. \rev{For these larger sources, } future investigations such as calculating the Pearson correlation coefficient of polarized intensity with the Galactic and extra-galactic emission, or examining gradients of the total intensity and their correlation with the total or gradients in the observed polarized intensity and Faraday structures \rev{could be useful. } 


\rev{However, even if the sources under study are small, or if some reduction in Galactic foregrounds can be made by masking out HII regions, or star-formation activity, there can be other unrelated foregrounds, such as the intervening intracluster medium.  Our approach to identifying likely foregrounds was subjective, and this must be put on a firmer, more quantitative basis.  Developments are necessary both for completely blind approaches, e.g., statistical correlations between the synchrotron and Faraday structures, as well as approaches based on radio galaxy models, as mentioned briefly below.}

\rev{Algorithms are also needed to more robustly identify when variations along the line of sight, (as opposed to e.g., local $n_eB$ enhancements from the intervening molecular disk of Cen~A)  dominate the local Faraday structure.  Here, we used the presence of coincident bending of the structure in the plane of the sky and in Faraday space in 3C40B, to argue for line of sight changes.  A more convincing case could be made for repeated structures, such as the suggestion of Faraday variations for ``corkscrew" tails \citep[Fig. 3 in][]{2015aska.confE.101J}. }

\rev{We note that automated algorithms, including the use of machine learning, and perhaps even artificial intelligence techniques, likely have an important role to play in the future of these studies.  This is especially important as the data volume increases dramatically.  A much more sophisticated understanding of the observational signatures of different realistic physical structures is a pre-requisite.  }


    \rev{\subsection{Visualization development}}\label{sec:vis}

A major part of the visualization challenge is choosing the appropriate resolution in Faraday space, as explained in Appendix A.  Techniques that utilize variable resolution, depending on the local signal:noise, should be explored. \rev{Scalable precision imaging  \citep{2023MNRAS.522.5558W}, based on the unconstrained Sparsity Averaging Reweighted Analysis (uSARA) optimization algorithm \citep{2023MNRAS.518..604T}, is one recent promising development.}

The visualization techniques for \PS~ cubes should \rev{also} be extended beyond those used here.  We utilized projections, movies from various angles, and small cross-sectional cuts, but other options \rev{need to be explored.  The powerful \emph{SlicerAstro} package \citep{2017A&C....19...45P},  provides a rich set of interactive options, including transparency, cuts through the cube with arbitrary planes, and signal:noise-based smoothing. It was developed for HI cubes, but can be used in its current form for Faraday cubes; some simple modifications would also improve this new utilization.}

Finally, there needs to be appropriate ways to exploit the information in full Faraday cubes, where each spatial pixel may contain information from multiple Faraday components. \rev{ Issues such as distinguishing primary from secondary peaks in the spectra, effective deconvolution techniques and the appropriate choice of the display beam resolution (see Appendix A), all need to be addressed.} 

\rev{\subsection{Science development}}\label{sec:scidev}
\rev{The most pressing need is for simulations, and radio source modeling in general, which includes the presence of a thermal plasma mixed with the synchrotron plasma and in the immediate surroundings.  Mixing needs to be explored both on macroscopic scales (e.g., synchrotron filaments embedded in a thermal plasma lobe), and on microscopic scales (e.g.,  mixed synchrotron/thermal jets).   The sophisticated 3D polarization simulations of relativistic MHD jets by \cite{2023arXiv231112363J},  visualized with \PS, are one such example. }

\rev{\cite{2023MNRAS.526.5418M} provide a recent example of the kind of external medium studies needed, looking at the interaction of AGN jets with a surrounding turbulent medium.  Another important case is magnetic draping, such as suggested by \cite{2019A&A...622A.209A} and \cite{LG2}, and explored theoretically by \cite{2011gcca.progE..23P}. A similar situation has been explored on Galactic scales, where  filaments which are moving with respect to the external gas, and can produce an arc-shaped morphology as the fields bend around the filaments \citep{LiKlein2019,Tahanietal2019, Tahanietal2022O, Tahanietal2022P}.  The signatures of all of these in the \PS~ cubes need to be explored.}

Another topic to be investigated is the use of other information from the emitting source to compare with the Faraday structures.  These could include total intensity structures, as well as spectral indices, polarization angles and depolarization variations. 
\vspace{0.15in}

\section{Conclusions}\label{sec:concl}
We have presented a simple visualization technique to create Pseudo-3D cubes of Faraday structure from pairs of polarized intensity and rotation measure maps, \rev{and provided a Python tool for this purpose.}  Using subjective, cartoon-based models, we have illustrated how to separate Faraday variations due to foregrounds unrelated to the synchrotron emitting source from those with which the source is mixed or interacting.   We have applied this to several different situations, and showed how information on local Faraday structures, 3D source structures, jet orientations and the internal magnetic field configuration of jets can be derived.  We find that there are many examples where these local effects are important, and thus provide new opportunities to study radio galaxy physics, including the interactions of the relativistic and surrounding thermal plasmas. \rev{It is important to understand that any models can be used to check for consistency, but do not represent unique interpretations of the observations.}

We identify a number of areas where additional work is needed to exploit the use of these Pseudo-3D cubes; \rev{these involve algorithm development to put the interpretations on a more quantitative basis, visualization development to allow more powerful exploration of the \PS~ cubes, and scientific development to illustrate in a more comprehensive way how thermal and relativistic plasmas could be inter-related.} 

We recommend that these \PS~ techniques be used both on past and future polarization mapping projects, to inform our understanding of the physics of radio galaxies and their interactions with the thermal plasma mediums in which they are embedded.

    \section*{Acknowledgements}
A.P. acknowledges the program ``Programade Investigadoras e Investigadores por M\`{e}xico” (CONACyT). Very Large Array data and support for WC comes from the National Radio Astronomy Observatory, which is a facility of the U.S. National Science Foundation operated under cooperative agreement by Associated Universities, Inc.. M.T. is supported by the Banting Fellowship (Natural Sciences and Engineering Research Council Canada) hosted at Stanford University and the Kavli Institute for Particle Astrophysics and Cosmology (KIPAC) Fellowship.   \rrev{Suggestions by the anonymous referee and the editor have led to significant improvements in the paper.}
The MeerKAT telescope is operated by the South African Radio Astronomy Observatory, which is a facility of the National Research Foundation, an agency of the South Africa Department of Science and Innovation. This research has made use of SAOImageDS9, developed by Smithsonian Astrophysical Observatory.

\section*{Data Availability}   The polarization maps used as input for the production of \PS~ are available as  noted in the citations for each source, from the individual originating observatories.  The \PS~ FITS cubes produced here will be made available by the authors upon reasonable request.

 \clearpage
 \bibliographystyle{mnras}
\bibliography{main2.bib}
 \clearpage
\section*{Appendix A: The Display Challenge}\label{sec:app}

\rev{As is well known,} restoring an image with a clean beam in normal radio interferometry \rev{frequently} obscures the accuracy, $\delta\theta$,\footnote{$\delta\theta = FWHM/(2\times S/N)$ where FWHM is the clean beam size and S/N is the signal to noise ratio.} of a feature's position. $\delta\theta$ is often less than even the pixel size. Attempts to reflect the actual \rev{positional} accuracy of features include maximum entropy reconstructions \citep{1977ITCom..26..351W}, \rev{and} adaptive smoothing \citep{2006MNRAS.368...65E} for X-rays. Multi-resolution clean algorithms \citep{2008ISTSP...2..793C} also have a variable effective resolution depending on  brightness  \rev{and the new \emph{uSARA} algorithm \citep{2023MNRAS.518..604T} holds great promise.} In this section, we use the term \emph{display beam} to designate what is used for imaging, to explicitly distinguish it from the FWHM of the synthesized beam.

\begin{figure}
    \centering
    \includegraphics[width=3.0in]{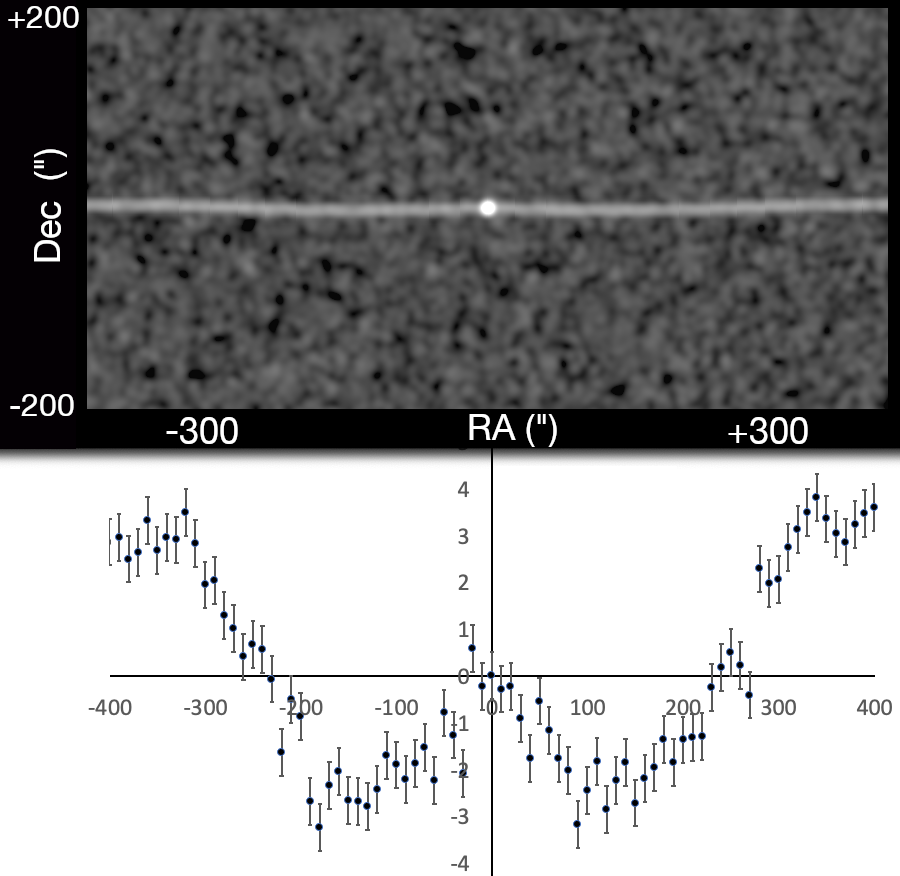}
   
    \caption{A simulated core and jets, with wiggles along the jets that are much less than its width, as described in the text. \rev{Barely visible in the greyscale (top), the wiggles are detected with high signal:noise in the Gaussian fits (bottom);} errors represent the statistical error in the mean positions. } 
    \label{fig:jet}
    \end{figure}

Figure \ref{fig:jet} illustrates the issue. The two dimensional picture  shows two jets with a signal:noise of 15;  they could be straight, or have wiggles that are \rev{either symmetric or} anti-symmetric. The  beamwidths of both the synthesized beam and the display beam are 10\arcsec, while the jet has \rev{transverse} oscillations  of only $\pm 3\arcsec$. In the 2D image,  some curvature is just visible,  but any possible symmetry or anti-symmetry is unclear. 
However,  Gaussians \rev{fit} to transverse profiles across the jets and a plot of the offset of their peaks ( Fig. \ref{fig:jet} bottom), \rev{show the} offsets and symmetry at very high significance. 

The same problem occurs in Faraday synthesis mapping, where  the accuracy $\delta\Phi$ of a component's peak Faraday depth can be much smaller than the Faraday beam width or even the spectrum sampling width.  
In standard practice to date, the Faraday ``display beam" is taken to be the width of the main peak in the Faraday amplitude spectrum \citep{Brentjens2005}.\footnote{ \cite{RudnickCotton} have shown that reliable information from the complex Faraday spectra can be revealed using the narrower width of the real component of the spectrum ('full resolution').} 
Fine-scale reliable variations in the peak of the Faraday spectrum, \rev{and in the 2D images (RA, $\Phi$) and (Dec, $\Phi$) can } be obscured if we use the nominal, \rev{and even the full resolution} beam. 

In this paper, we made a \rev{somewhat} arbitrary choice of \rev{the display beam width, implemented with a Faraday smoothing width, $\Phi_{sm}$, of typically 3-5 $rad/m^2$}; \rev{we sought to} maximize the visibility of the variations without creating artifacts due to oversampling. This needs more quantitative analysis and a robust discussion in the community about how to reliably represent the full available information.

\begin{figure}
    \centering
    \includegraphics[width=3.in]{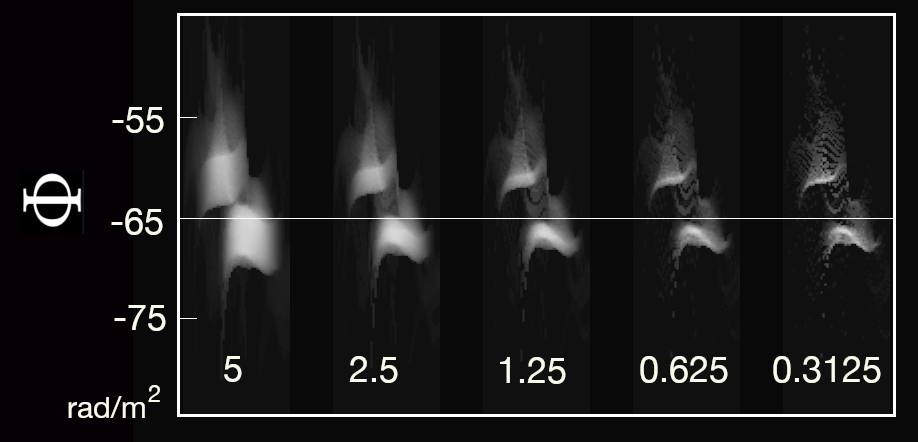}

    \caption{Five juxtaposed (RA, $\Phi$) maps of the NE lobe of Cen A for different display beam sizes, listed in \radmm.}  
    \label{fig:smooth}
    \end{figure}

    The effects of different Faraday display beam choices are shown in Fig. \ref{fig:smooth} for the northeastern lobe of the inner double of Cen~A, \rev{(Case 2)}. The bimodal Faraday structure of the lobe, with a separation of $\lessapprox 5$~\radmm, would be invisible at the nominal Faraday resolution of $\sim 60$~\radmm~ \citep{Centaurus}. Here, even with a display beam as large as 5~\radmm, it is easily visible.  The detailed shape of the Faraday structure becomes more apparent with the smaller display beams.  At the same time, in the regions of lower brightness, the faint emission breaks up into spurious narrow bands; the structure on these very fine scales is not reliable.  \rrev{Such features appear in a variety of situations where oversampling is present.} 

    \rrev{This is related to the broader issue of how to identify spurious structures. In reference to their Fig. 19, \cite{RudnickCotton} note, ``... when significant Faraday variations occur within a spatial beam, spurious structures can appear in the tomography images [images at different Faraday depths, as in the pseudo-3D cubes here]."  Some signatures of spurious features include the parallel bands mentioned above, spatially unresolved jumps in Faraday depth, and the presence of significant depolarization, which indicates the presence of multiple Faraday components at the same location.} Again, these visual impressions need to be investigated quantitatively, and algorithms for choosing appropriate display beam widths \rrev{and recognizing spurious structures need to be} developed.
    
\clearpage
\section*{Appendix B: Pseudo-3D script and usage notes}\label{sec:code}

We provide a Python script, \emph{generate\_pseudo3D.py}, to facilitate the creation of pseudo-3D cubes.  \rev{It is} available at \href{https://github.com/candersoncsiro/rmsynth3d}{~ GITHUB}. \rev{The script} takes pairs of matched polarized intensity, Faraday depth maps and creates a pseudo-3D cube, \PS.  Users should first prepare the input maps to have the desired angular resolution, and optionally to mask out low polarized intensity regions.   For \rev{spatially} elongated structures it will also be advantageous to rotate the maps so that the major axis is  either horizontal or vertical.  

If a \PS cube is desired for only a smaller cross section of the source, as in Fig. \ref{fig:M87slices}, then the simplest approach is to  make an input pair of (PI,$\Phi$) images of only that area.  Alternatively, a portion of the output \PS~ cube can be extracted, as needed, but this must be done before visualization of the cube from different angles. 

The required input parameters for \emph{generate\_pseudo3D.py} are the minimum and maximum Faraday depths to be included, the pixel size in Faraday depth, and a Faraday smoothing width. We recommend that the boundaries in Faraday depth be somewhat larger than the range of significant depths in the image, and \rev{that} the pixel size and smoothing width \rev{be} guided by the considerations in Appendix A.

 \rev{For visualization, we used \emph{SAOimage DS9} in 3D Frame mode to view the cube from different directions. We used } Average Intensity Projection (AIP) rendering, and then created movies using 1$\degree$ viewing angle increments \rev{back and forth} over \rev{90}$\degree$, typically around the major axis. Zooming in on the $\Phi$ axis is often useful, and is accomplished in \emph{DS9} using the 3D Frame parameter \emph{Z~Axis~Scale}. \rev{We also strongly recommend investigations using the interactive package \href{https://github.com/Punzo/SlicerAstro/wiki}{\emph{SlicerAstro}}, \citep{2017A&C....19...45P}. It has a steep learning curve but offers a large number of additional tools, such as transparency and slicing of the cube.}

We note again that \rev{$\mathds{F}$} cubes created from the full Faraday spectrum at each pixel are \rev{similar to}  ``pseudo-3D" cubes in that the Faraday axis may or may not correspond to the actual third spatial dimension.  Movies of \rev{$\mathds{F}$ cubes} can \rev{also} be useful, but are \rev{at a much lower resolution than \PS~ cubes} and subject to confusion from emission away from the main peak. 

\end{document}